\documentclass{article}

\usepackage{arxiv}

\usepackage[utf8]{inputenc}
\usepackage[T1]{fontenc}
\usepackage[english]{babel}

\usepackage{microtype}
\usepackage{pdflscape}
\usepackage{float}

\usepackage{amsmath, amsfonts, amssymb}
\usepackage{nicefrac}
\usepackage{algorithm}
\usepackage{algpseudocode}

\usepackage{graphicx}
\graphicspath{{./figures/}}
\usepackage{subcaption}
\usepackage{standalone}
\usepackage{tikz}
\usetikzlibrary{
  shapes.geometric,
  shapes.symbols,
  arrows.meta,
  positioning,
  fit,
  calc,
  backgrounds
}
\pgfdeclarelayer{background}
\pgfsetlayers{background,main}

\usepackage{booktabs}
\usepackage{tabularx}
\usepackage[table]{xcolor}
\usepackage{csvsimple}

\usepackage{hyperref}
\usepackage{url}
\usepackage{natbib}

\usepackage{enumitem}
\usepackage{ulem}
\usepackage{verbatim}

\usepackage{tcolorbox}
\usepackage{mdframed}

\usepackage{listings}
\lstdefinestyle{mystyle}{
    basicstyle=\ttfamily\tiny,
    breaklines=true,
    breakatwhitespace=true,
    columns=fullflexible,
    keepspaces=true,
}

\usepackage{todonotes}

\definecolor{v0}{HTML}{31688e}
\definecolor{v1}{HTML}{90d743}
\definecolor{v2}{HTML}{443983}
\definecolor{v3}{HTML}{fde725}
\definecolor{v4}{HTML}{35b779}   % v4 redéfini une seule fois
\definecolor{c0}{HTML}{20a386}
\definecolor{c1}{HTML}{238a8d}
\definecolor{c2}{HTML}{2d718e}
\definecolor{c3}{HTML}{33638d}
\definecolor{c4}{HTML}{404688}
\definecolor{c5}{HTML}{482576}
\definecolor{viridisBlue}{HTML}{31688E}

\tikzset{
    startstop/.style={
        rectangle, rounded corners,
        minimum width=3cm, minimum height=1cm,
        text centered, draw=black, fill=red!30
    },
    process/.style={
        rectangle,
        minimum width=3cm, minimum height=1cm,
        text centered, draw=black, fill=blue!30,
        text width=5cm
    },
    process_/.style={
        rectangle,
        minimum width=3cm, minimum height=1cm,
        text centered, draw=black, fill=blue!10,
        text width=9cm
    },
    decision/.style={
        diamond, aspect=2, draw, fill=green!30,
        text width=4em, text centered, inner sep=0pt
    },
    arrow/.style={-{Latex}, thick}
}
\title{Spatio-Temporal Modelling of Electric Vehicle Charging Demand}

\author{%
  \begin{tabular}{cc}
    \textbf{Kaoutar Bouaachra} & \textbf{Yvenn Amara-Ouali} \\[0.2em]
    École Polytechnique de Paris & EDF Lab Paris-Saclay \\
    Palaiseau, France & Palaiseau, France \\
    \texttt{kaoutar.bouaachra@polytechnique.edu} & \texttt{yvenn.amara-ouali@edf.fr} \\[1.8em]
    \textbf{Yannig Goude} & \textbf{Raphaël Lachieze-Rey} \\[0.2em]
    EDF Lab Paris-Saclay & INRIA Paris, DYOGENE/MATHNET Team \\
    Palaiseau, France & Paris, France \\
    \texttt{yannig.goude@edf.fr} & \texttt{raphael.lachieze-rey@math.cnrs.fr} \\
  \end{tabular}
}

\begin{document}
\maketitle
\begin{abstract}
Accurate forecasting of electric vehicle~(EV) charging demand is critical 
for grid management and infrastructure planning. Yet the field continues to 
rely on legacy benchmarks; such as the Palo Alto (2020) dataset; that fail 
to reflect the scale and behavioral diversity of modern charging networks. 
To address this, we introduce a novel large-scale longitudinal dataset 
collected across Scotland (2022--2025), which release it as an open 
benchmark for the community.

Building on this dataset, we formulate EV charging demand as a 
spatio-temporal latent Gaussian field and perform approximate Bayesian 
inference via Integrated Nested Laplace Approximation~(INLA). The resulting 
model jointly captures spatial dependence, temporal dynamics, and covariate 
effects within a unified probabilistic framework. On station-level 
forecasting tasks, our approach achieves competitive predictive accuracy 
against machine learning baselines, while additionally providing principled 
uncertainty quantification and interpretable spatial and temporal 
decompositions properties that are essential for risk-aware infrastructure 
planning.
\keywords{ Electric vehicles \and Spatio-temporal modelling \and Charging demand \and Bayesian statistics \and integrated nested Laplace approximation (INLA)\and Machine learning.}
\end{abstract}

\section{Introduction}
The rapid growth of electric vehicles (EVs) is a key lever of global 
decarbonization strategies \citep{IEA2023}. As EV penetration increases, 
predicting charging demand has shifted from a long-term planning concern to 
an operational necessity for grid management and infrastructure development 
\citep{GNANN2018314}. Accurate modeling is essential not only for balancing 
aggregate power supply, but also for mitigating local grid congestion and 
ensuring user satisfaction \citep{NEAIMEH2015688}. Forecasting EV demand 
remains challenging due to its dual nature: it evolves simultaneously across 
high-frequency temporal cycles and heterogeneous spatial networks.

Modeling EV charging demand at scale presents two main challenges. First, 
accurate predictions require capturing localized spatial spillover effects 
alongside multi-scale temporal dynamics. Second, as the number of spatial 
locations and time steps grows, standard Bayesian inference methods such as 
Markov Chain Monte Carlo (MCMC) become computationally prohibitive. The goal is therefore to develop a framework 
that is both statistically principled and scalable to operational datasets.

\noindent\textbf{Related Works.}The modeling of EV charging demand has evolved from purely temporal approaches 
toward increasingly rich spatio-temporal frameworks. Early studies relied on 
aggregate time-series models such as ARIMA or Seasonal Holt--Winters methods 
to predict charging demand at the grid level \citep{5356176, RICHARDSON2013247, en14051487}. While effective for long-term planning, these 
approaches assume spatial homogeneity and fail to capture local congestion 
effects and distribution network constraints.

To account for spatial heterogeneity, agent-based models (ABMs) have been 
proposed to represent individual vehicle trajectories and charging behaviors 
\citep{FENG2026106864}. Although granular, these models are computationally 
intensive, sensitive to behavioral assumptions, and offer limited uncertainty 
quantification. More recently, machine learning approaches including LSTMs 
and graph convolutional networks have shown strong predictive performance for 
spatio-temporal EV demand forecasting \citep{Yu_2018, wu2019graphwavenetdeepspatialtemporal, TIAN2025125174}.

In this work, we adopt a complementary perspective based on Latent Gaussian 
Models (LGMs). Leveraging the Integrated Nested Laplace Approximation (INLA) 
framework \citep{RUE20073177}, our approach combines scalable Bayesian 
inference, principled uncertainty quantification, and interpretability. This 
methodology has proven effective in environmental and geostatistical modeling 
but has not yet been applied to EV charging demand forecasting. Compared to 
MCMC, INLA enables efficient large-scale spatio-temporal inference, making it 
well suited for operational decision-making.
-
Beyond modeling approaches, several surveys have reviewed the landscape of 
publicly available EV charging datasets \citep{en14082233, CALEARO2021111518, 
10670452}. \citep{en14082233} reviewed more than 860 open repositories 
and identified around 60 relevant datasets covering charging sessions across 
multiple countries, while \citep{CALEARO2021111518} and \citep{10670452} provide 
broader overviews of data sources used for EV integration and forecasting 
studies. To complement these existing resources, this work introduces a 
large-scale longitudinal dataset of EV charging sessions in Scotland, spanning 
October 2022 to April 2025, which reflects recent trends in EV adoption and 
infrastructure deployment.

\noindent\textbf{Main Contributions.} This paper makes two main contributions. First, we introduce a longitudinal 
dataset of EV charging demand in Scotland (October 2022 -- April 2025), 
together with a reproducible preprocessing pipeline that integrates raw 
charging transactions with fine-scale geographic information. Second, we 
formulate EV charging demand as a spatio-temporal latent Gaussian model and 
perform approximate Bayesian inference using INLA, jointly capturing spatial 
dependence, temporal dynamics, and covariate effects within a unified 
probabilistic framework.

\noindent\textbf{Overview.}
The remainder of the paper is organized as follows. 
Section~\ref{sec:data} presents the dataset, the preprocessing pipeline, 
and an exploratory analysis of spatio-temporal charging patterns over the 
2022--2025 period. Section~\ref{Problemdes} introduces the spatio-temporal 
modeling framework and the evaluation protocol. Section~\ref{results} reports 
the empirical results and compares the proposed models against station-level 
baselines. 

% ─── Section 2 ─────────────────────────────────────────────────────────────────
\section{EV Charging Dataset}
\label{sec:data}
 
\subsection{Data Collection and Preprocessing}
The integrity of spatio-temporal modeling for Electric Vehicle (EV) demand is fundamentally dependent on the quality of the underlying charging records. This section details our data provenance and the multi-stage curation pipeline developed to ensure a physically plausible and consistent dataset.

\noindent \textbf{Data Acquisition:} 
The dataset used in this study consists of EV charging session records 
from Scotland, sourced from the ChargePlace Scotland open-access 
repository \citep{ChargePlaceScotland}. We selected this source for 
three reasons: the data is publicly available and auditable, it covers 
a recent period (October 2022 to April 2025) that captures current 
trends in EV adoption, and this temporal range has not previously been 
studied in the spatio-temporal modeling literature. We enriched the raw 
session logs with auxiliary metadata from \citep{wevj15120570}, mapping 
charge-point identifiers to connector specifications and postcodes. 
Geographic boundaries from Open Data Scotland \citep{OpenDataScotland} 
were added to enable territorial indexing at multiple spatial levels.

\noindent \textbf{Data Enrichment:} 
The raw data consists of monthly session logs. For each record, we 
extracted start and end timestamps, session duration, energy consumed 
($kWh$), and the local authority identifier. By joining with the 
\citep{wevj15120570} reference set, we added charger power ratings 
(slow, fast, and rapid) and postcodes. Each unique charge-point 
identification code was treated as a distinct physical asset.

\noindent \textbf{Data Curation:} 
We removed sessions with zero duration, zero energy transfer, or null 
entries, as these typically correspond to failed connection handshakes 
rather than actual charging events. Sessions exceeding the maximum 
usable battery capacity of the UK mass-market EV fleet were discarded 
\citep{battery_capacity}. Sessions whose billed cost exceeds the 
maximum daily tariff, including overstay penalties, were also removed 
\citep{tariff_fee}.

\noindent \textbf{Standardization:} 
The raw data contained inconsistent date-time formats 
(\texttt{DD/MM/YYYY} vs.\ \texttt{MM/DD/YYYY}), which required custom 
parsing routines. We identified and removed specific time windows with 
systematic reporting failures, notably in October 2022, March and June 
2023, and early 2024, to avoid artificial gaps in the time series. 
Local authority names were standardized for categorical consistency. 
The final cleaned dataset is stored as \texttt{master.csv} and serves 
as the input for all downstream analyses. It was assumed that each unique charge-point identification code represented a distinct charging point.

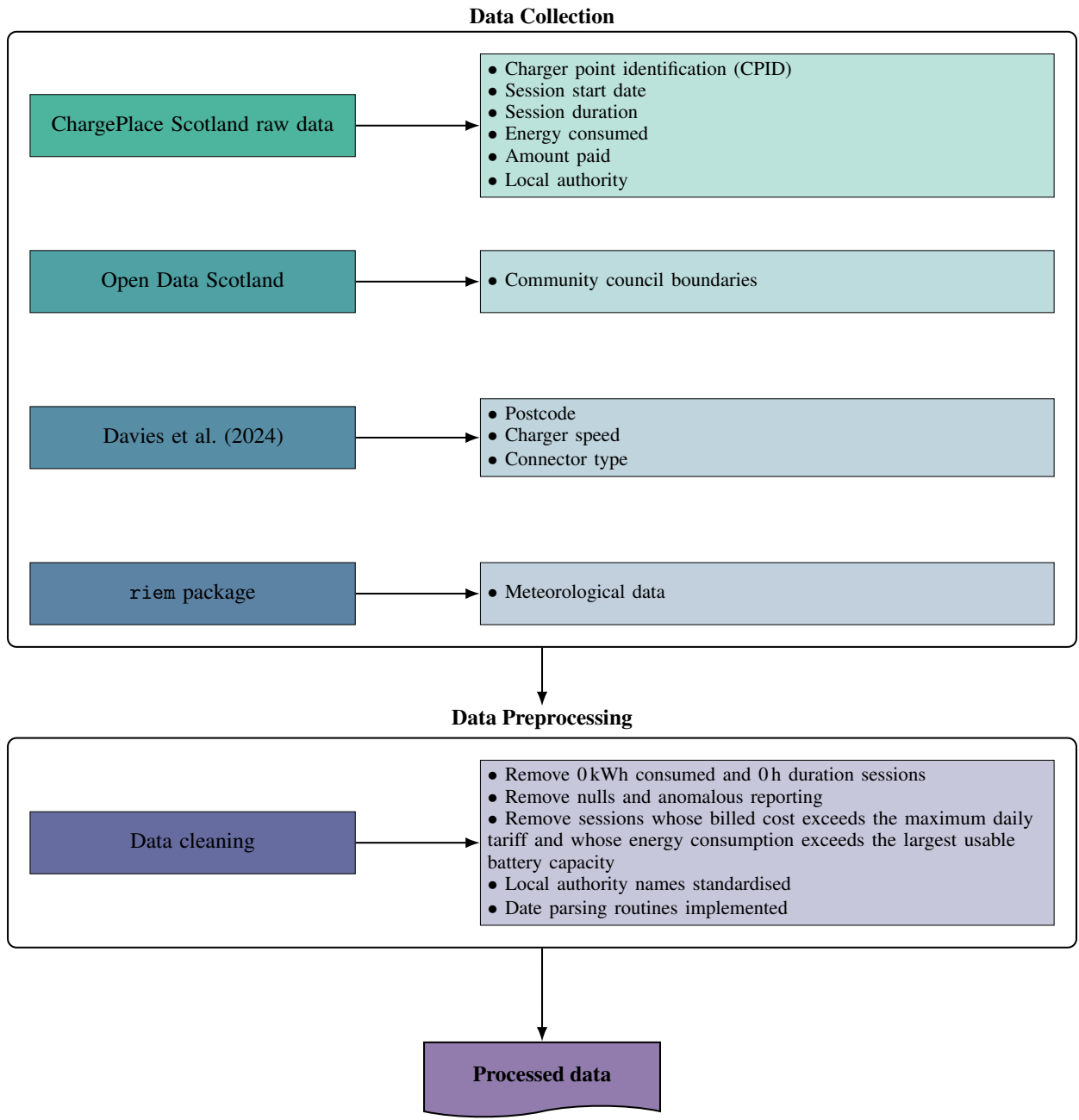
\begin{figure}[ht]
    \centering
    \begin{tikzpicture}[node distance=1.5cm and 2cm]

% --------------------
% Data collection
% --------------------
\node (ChargePlace) [process, fill=c0!80, draw=black!100!black] {ChargePlace Scotland raw data};
\node (OpenData)    [process,fill=c1!80, draw=black!100!black, below=of ChargePlace] {Open Data Scotland};
\node (Davis)       [process, fill=c2!80, draw=black!100!black,below=of OpenData] {Davies et al.\ (2024)};
\node (Meteo)       [process,fill=c3!80, draw=black!100!black, below=of Davis] {\texttt{riem} package };

\node (ChargePlace_) [process_,fill=c0!30, draw=black!100!black, right=of ChargePlace, align=left] {\small
$\bullet$ Charger point identification (CPID)\\
$\bullet$ Session start date\\
$\bullet$ Session duration\\
$\bullet$ Energy consumed\\
$\bullet$ Amount paid\\
$\bullet$ Local authority
};

\node (OpenData_) [process_,,fill=c1!30, draw=black!100!black, below=of ChargePlace_, right=of OpenData, align=left] {\small
$\bullet$ Community council boundaries
};

\node (Davis_) [process_,fill=c2!30, draw=black!100!black, below=of OpenData_, right=of Davis, align=left] {\small
$\bullet$ Postcode\\
$\bullet$ Charger speed\\
$\bullet$ Connector type
};

\node (Meteo_) [process_, fill=c3!30, draw=black!100!black, below=of Davis_, right=of Meteo, align=left] {\small
$\bullet$ Meteorological data
};

\node[
  draw, rounded corners, thick, inner sep=10pt,
  fit=(ChargePlace) (OpenData) (Davis) (Meteo)
      (ChargePlace_) (OpenData_) (Davis_) (Meteo_),
  label={[font=\bfseries]above:Data Collection}
] (Data collection) {};

\draw[arrow] (ChargePlace) -- (ChargePlace_);
\draw[arrow] (OpenData)    -- (OpenData_);
\draw[arrow] (Davis)       -- (Davis_);
\draw[arrow] (Meteo)       -- (Meteo_);

% --------------------
% Data preprocessing
% --------------------
\node (Data cleaning) [process, fill=c4!80, draw=black!100!black, below=3cm of Meteo] {Data cleaning};

\node (Data cleaning_) [process_,fill=c4!30, draw=black!100!black,  below=3cm of Meteo_, right=of Data cleaning, align=left] {\small
$\bullet$ Remove 0\,kWh consumed and 0\,h duration sessions\\
$\bullet$ Remove nulls and anomalous reporting\\
$\bullet$ Remove sessions whose billed cost exceeds the maximum daily tariff 
and whose energy consumption exceeds the largest usable battery capacity\\
$\bullet$ Local authority names standardised\\
$\bullet$ Date parsing routines implemented
};

\node[
  draw, rounded corners, thick, inner sep=10pt,
  fit=(Data cleaning) (Data cleaning_),
  label={[font=\bfseries]above:Data Preprocessing}
] (Data preprocessing) {};

\draw[arrow] (Data cleaning) -- (Data cleaning_);
\draw[arrow] (Data collection.south) -- ($(Data preprocessing.north)+(0,0.5cm)$);

% --------------------
% Output
% --------------------
\node (Processed) [
  tape, draw, thick,
  tape bend top=none,fill=c5!60, draw=black!100!black,
  font=\bfseries,
  minimum width=3.8cm,
  minimum height=1.2cm,
  align=center,
  below=of Data preprocessing
] {Processed data};

\draw[arrow] (Data preprocessing.south) -- (Processed);

\end{tikzpicture} 
    \caption{\small  Overview of data collection and preprocessing steps.}
    \label{fig:data_collection_preprocessing}
\end{figure}

The comprehensive workflow, ranging from the initial data collection from the ChargePlace Scotland (CPS) network to the final feature engineering phase, is illustrated in Figure~\ref{fig:data_collection_preprocessing}.

\newpage
\subsection{Exploratory Data Analysis }
\label{sec:eda}

Before developing spatio-temporal models, we perform an exploratory 
analysis of the dataset to better understand the charging patterns and 
identify the main drivers of demand. The analysis covers three aspects 
of the Scottish charging network:

\begin{itemize}
    \item \textbf{Temporal patterns:} We examine daily and weekly usage 
    patterns to identify peak demand periods and recurring seasonalities.
    
    \item \textbf{Tariff effects:} We assess how the transition from free 
    to paid access affected session volumes across different regions.
    
    \item \textbf{Charger types:} We compare usage profiles between AC 
    and Rapid/Ultra-Rapid chargers to understand how infrastructure 
    characteristics shape demand.
\end{itemize}

\noindent \textbf{Temporal Patterns:} 
Charging activity shows clear differences between weekdays and weekends, 
reflecting the influence of work and leisure schedules on usage behavior.
\begin{figure}[h]
    \centering
    \includegraphics[width=0.9\linewidth]{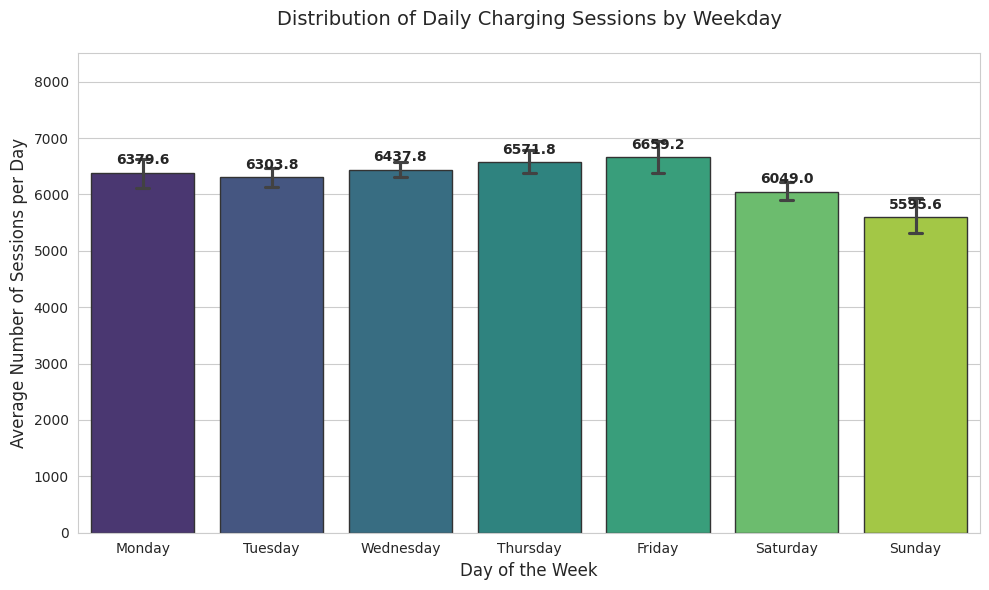}
    \caption{\small  \textbf{Average Daily Session Volume by Weekday.} Bars represent the mean number of charging sessions per day, while error bars denote the 95\% confidence interval (CI). This distribution highlights the weekly seasonality and the stability of charging demand.}
    \label{fig:distribution of daily charging sessions per weekday}
\end{figure}
As shown in Figure~\ref{fig:distribution of daily charging sessions per 
weekday}, session counts are higher on weekdays than weekends, with 
Friday recording the highest average demand. Bootstrapped confidence 
intervals reveal that Friday, Sunday, and Monday exhibit higher 
variability, likely because these days mix professional and leisure 
travel patterns. Saturday, by contrast, shows the lowest variability 
despite a lower mean volume, suggesting more regular and predictable 
charging behavior. These differences justify including the day of the 
week as a categorical covariate in the model.

\begin{figure}[h]
    \centering
    \includegraphics[width=0.5\textwidth]{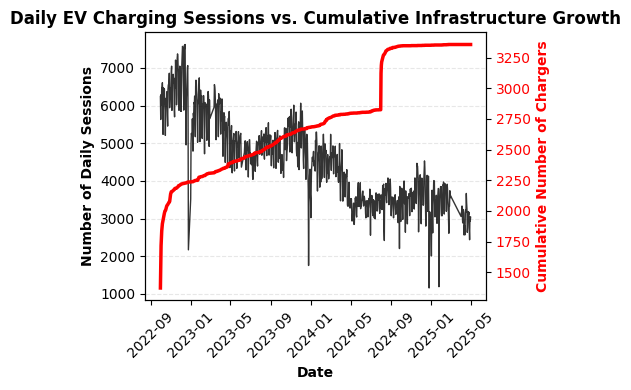}
    \caption{\small Evolution of the daily number of charging sessions and the number of public chargers recorded on the ChargePlace Scotland network over the entire dataset period.}

    \label{fig:sessions_vs_chargers}
\end{figure}

\noindent\textbf{Tariff Effects:} 
Figure~\ref{fig:sessions_vs_chargers} shows that while the number of 
available chargers grew steadily over the observation period, the total 
number of sessions declined. This apparent paradox is largely explained 
by tariff introductions: between October 2022 and April 2025, twelve 
local authorities transitioned from free to paid access.

\begin{figure}[h]
    \centering
    \begin{minipage}{0.32\textwidth}
        \centering
        \includegraphics[width=\linewidth]{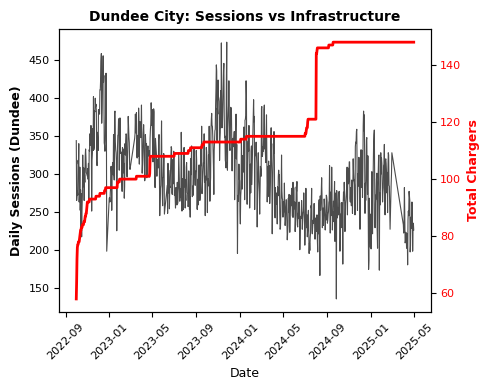}
        \caption{ \small Dundee City}
        \label{fig:dundee}
    \end{minipage}\hfill
    \begin{minipage}{0.32\textwidth}
        \centering
        \includegraphics[width=\linewidth]{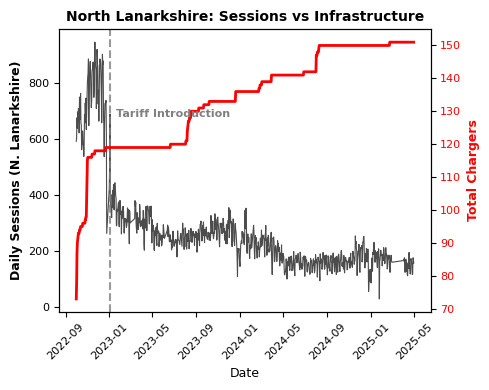}
        \caption{\small  North Lanarkshire}
        \label{fig:north_lanarkshire}
    \end{minipage}\hfill
    \begin{minipage}{0.32\textwidth}
        \centering
        \includegraphics[width=\linewidth]{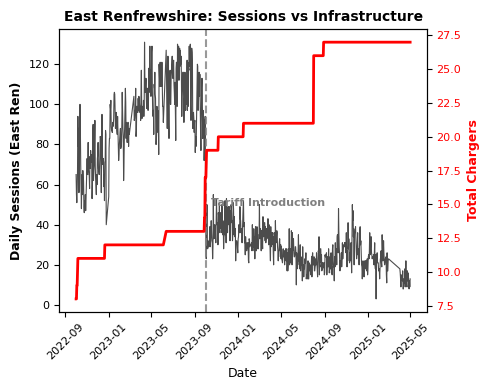}
        \caption{\small  East Renfrewshire}
        \label{fig:east_renfrewshire}
    \end{minipage}
\end{figure}

To illustrate this effect, we compare three regions with different 
pricing histories. In Dundee City, where tariffs were already in place 
before our observation window, session volumes remain stable 
(Figure~\ref{fig:dundee}). In North Lanarkshire and East Renfrewshire 
(Figures~\ref{fig:north_lanarkshire} and \ref{fig:east_renfrewshire}), 
session counts dropped sharply immediately after tariff introduction, 
as marked by the vertical dashed lines. This suggests that the 
introduction of costs had a strong and lasting effect on demand, 
particularly in regions where charging had previously been free. These results suggest that charging status (free vs.\ paid) should be 
included as a time-varying covariate in the model, as ignoring tariff 
changes would introduce systematic bias in the predictions.


\noindent\textbf{Public and Private Charge Points:} 
The \texttt{is\_public\_access} attribute distinguishes between public 
and private charge points. Public charge points show higher and more 
regular session volumes, though they remain sensitive to tariff 
changes as discussed above. Private charge points, which serve 
workplaces or vehicle fleets, record lower volumes with more irregular 
activity (Figure~\ref{publicprivate}). We therefore include 
accessibility type as a categorical covariate in the model.

\begin{figure}[h!]
    \centering
    \includegraphics[width=0.8\linewidth]{figures/public_vs_private.png}
    \caption{\small Daily number of charging sessions for both Public and Private access points over the observed period.}
    \label{publicprivate}
\end{figure}

\begin{figure} [h!]
    \centering
    \includegraphics[width=0.6\linewidth]{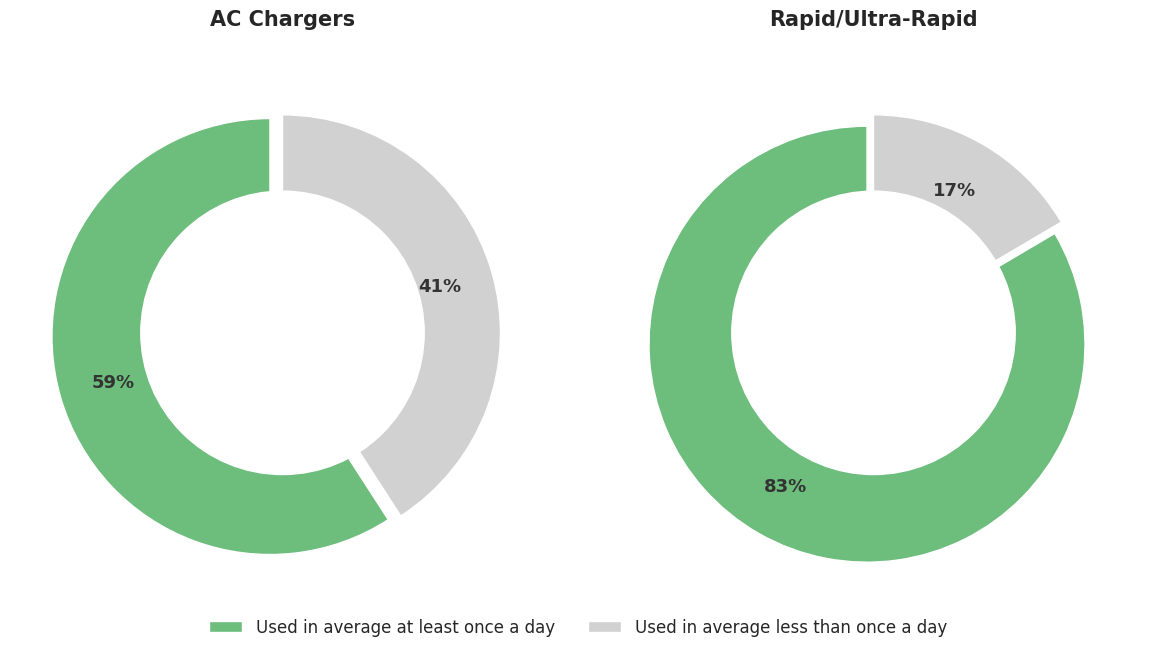}
    \caption{\small \textbf{Average daily activity rates by charger category.} In total, 59\% of AC chargers (left donut chart) experience at least one daily session, 
     In contrast, 
    83\% of rapid/ultra-rapid chargers (right donut chart) experience at least one daily session.} 
    \label{fig:chargersuse}
\end{figure}

\noindent\textbf{Charger Types and Utilization:} 
Figure~\ref{fig:chargersuse} shows that Rapid and Ultra-Rapid chargers 
are used on 83\% of days on average, compared to 59\% for AC chargers. 
This means that 41\% of days record no sessions at AC charge points, 
against only 17\% for Rapid ones. After tariff introduction, session 
volumes dropped by 46.6\% at Rapid chargers and by 66.1\% at AC 
chargers, suggesting that AC demand is more price-sensitive. These 
differences in usage intensity and price sensitivity justify including 
connector type as a categorical covariate in the model.
\newpage
\subsection{Focus on Central Glasgow for Spatio-Temporal Demand Modeling}

\begin{figure}[H]
    \centering
    \includegraphics[width=0.9\linewidth, trim={0 0 0 2cm}, clip]
    {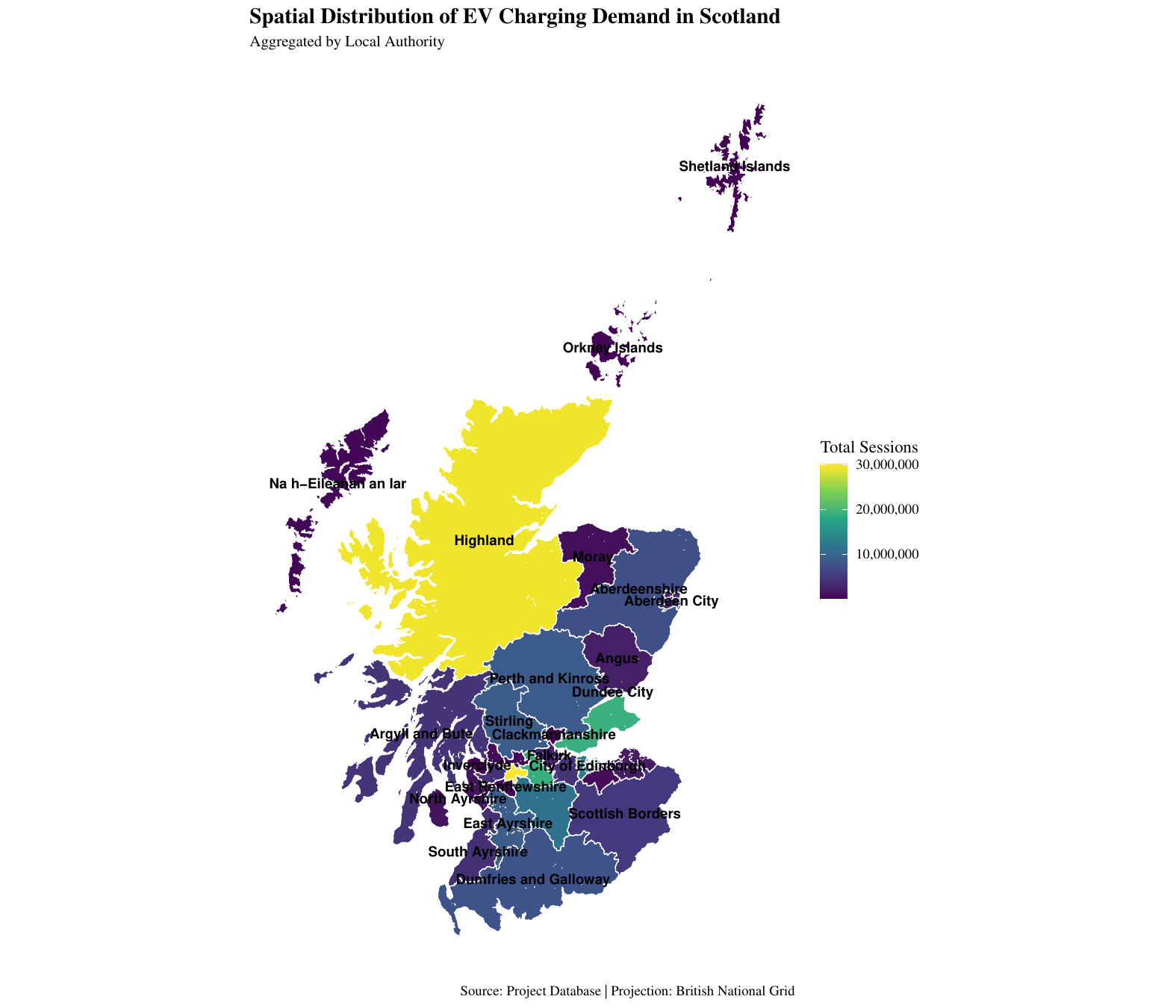}
    \caption{\small Spatial distribution of EV charging demand in 
    Scotland aggregated by local authority.}
    \label{fig:carte_ev}
\end{figure}

Figure~\ref{fig:carte_ev} shows the cumulative session volumes across 
Scottish local authorities. Glasgow City records the highest session 
density in the network and was therefore selected as the study area 
for the spatio-temporal benchmark.

\begin{figure}[H]
    \centering
    \begin{subfigure}[b]{0.48\textwidth}
        \centering
        \includegraphics[width=\linewidth]
        {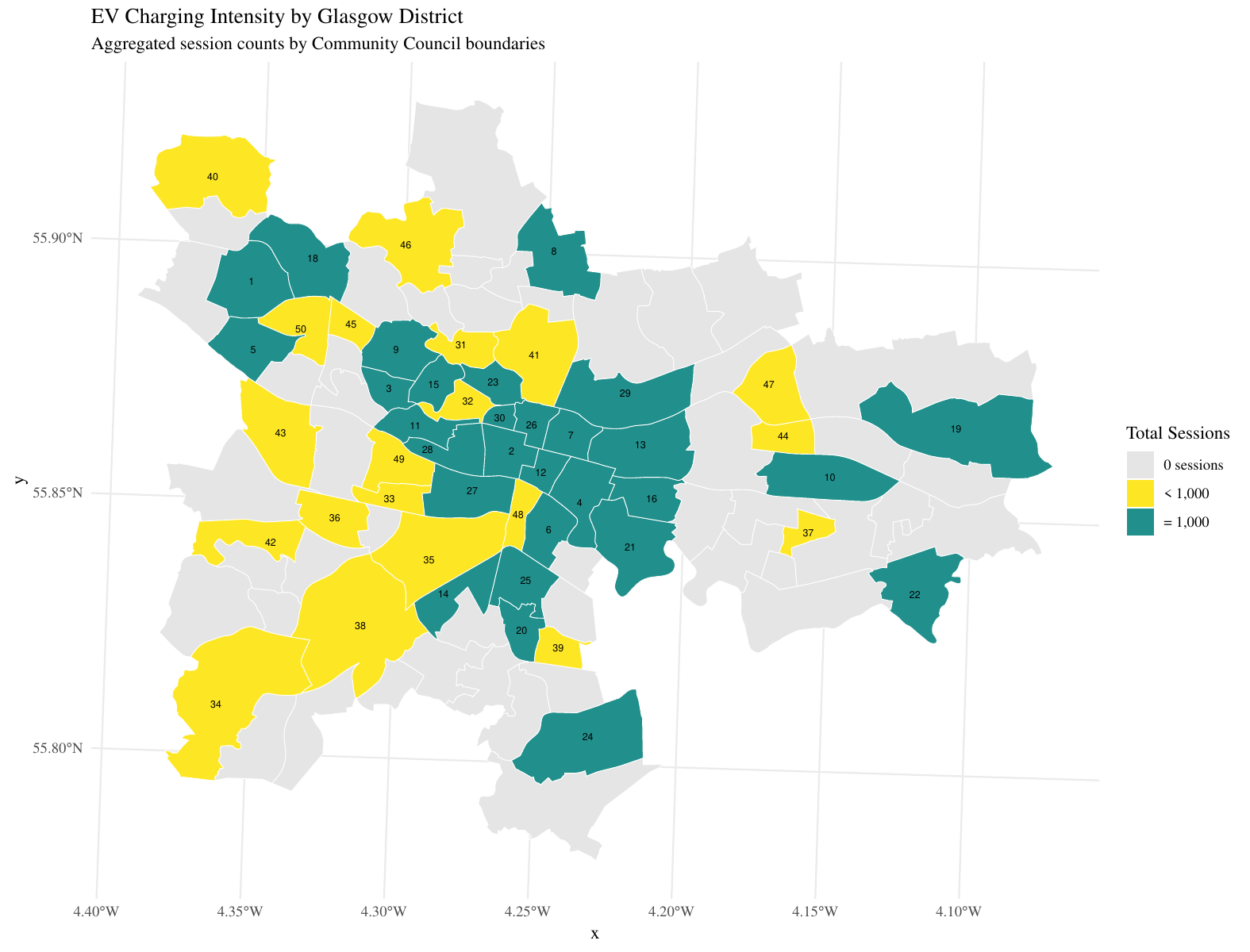}
        \caption{\small Geospatial distribution of charging activity in 
        Glasgow}
        \label{fig:study_area_map}
    \end{subfigure}
    \hfill
    \begin{subfigure}[b]{0.48\textwidth}
        \centering
        \includegraphics[width=\linewidth]
        {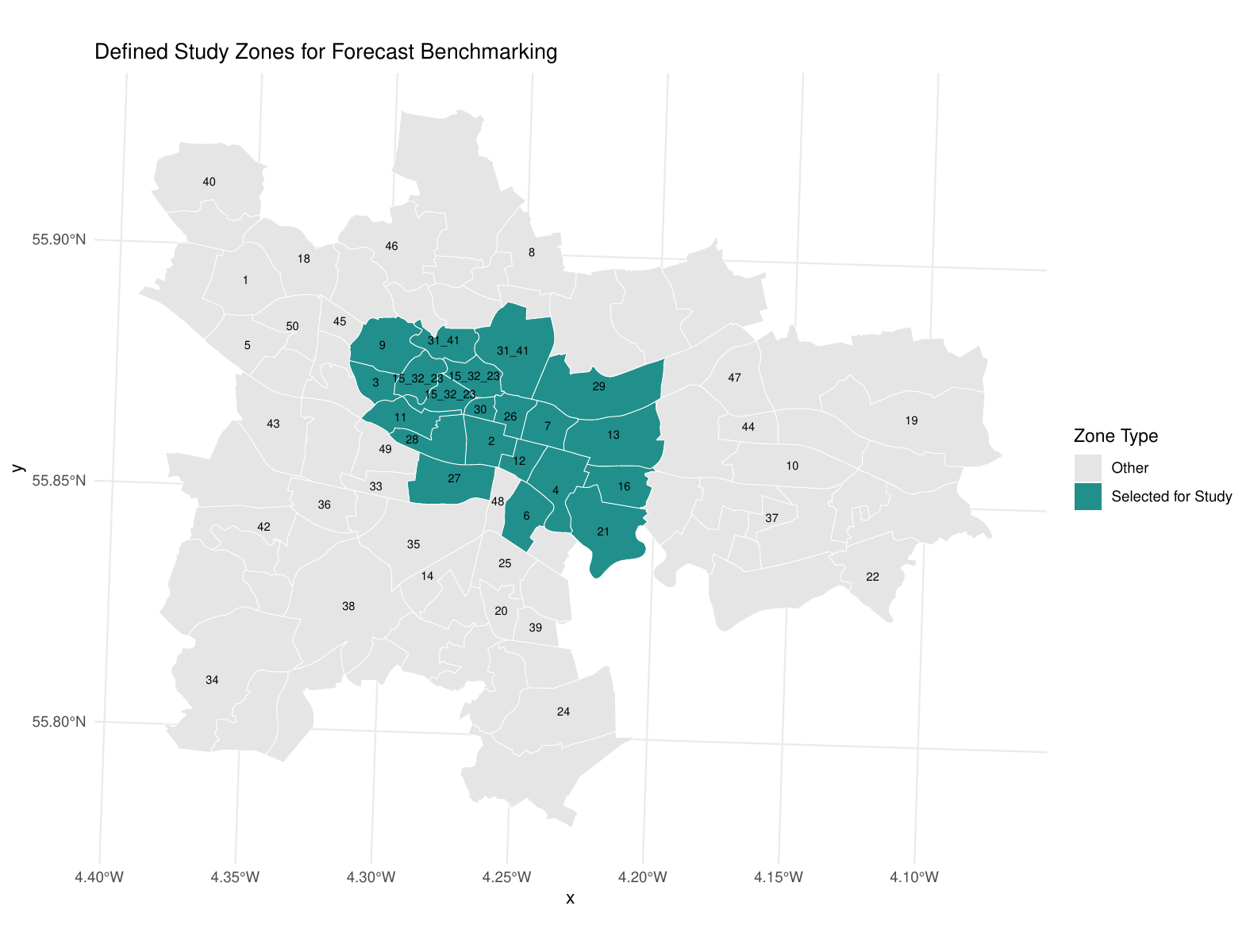}
        \caption{\small Selected zones for spatio-temporal demand 
        modeling in central Glasgow.}
        \label{fig:defined_study_zones}
    \end{subfigure}
    \caption{\small Spatial analysis of EV charging demand in Glasgow.}
    \label{fig:glasgow_maps}
\end{figure}

Figure~\ref{fig:study_area_map} shows the distribution of session 
volumes across Glasgow districts. High-activity zones ($>1{,}000$ 
sessions) are shown in green, low-activity zones ($<1{,}000$ sessions) 
in yellow, and areas with no data in Gray. The analysis is restricted 
to the urban core, where charging activity is sufficiently dense for 
spatio-temporal modeling.

We removed CPIDs with too few observations or persistent recording 
anomalies: CPIDs 62201, 62202, 62203, 62266, 62261, 50433, and 62123 
were excluded. The final subset, illustrated in 
Figure~\ref{fig:defined_study_zones}, covers 18 neighborhoods and 96 
CPIDs. Table~\ref{tab:dataset_stats} summarizes the key statistics of 
this dataset.
\begin{table}[h!]
\centering
\small
\caption{Summary statistics of the Glasgow EV charging dataset 
(October 2022 -- April 2025).}
\label{tab:dataset_stats}
\begin{tabular}{lc}
\toprule
\textbf{Property} & \textbf{Value} \\
\midrule
\multicolumn{2}{l}{\textit{Dataset scope}} \\
Observation period         & Oct 2022 -- Apr 2025 \\
Number of days             & 879 \\
Number of CPIDs            & 96 \\
Number of neighbourhoods   & 18 \\
Total session-day records  & 43,807 \\
\midrule
\multicolumn{2}{l}{\textit{Charging sessions (daily, per CPID)}} \\
Total sessions             & 104,041 \\
Mean                       & 2.37 \\
Median                     & 2 \\
Std.\ deviation            & 2.01 \\
Maximum                    & 32 \\
\midrule
\multicolumn{2}{l}{\textit{Infrastructure}} \\
AC chargers                & 92 (95.8\%) \\
Rapid chargers             & 4 \ \ (4.2\%) \\
Public charge points       & 86 (89.6\%) \\
Private charge points      & 10 (10.4\%) \\
Mean sessions -- AC        & 2.22 \\
Mean sessions -- Rapid     & 4.87 \\
\midrule
\multicolumn{2}{l}{\textit{Spatial extent}} \\
Longitude range            & $[-4.313,\ -4.210]$ \\
Latitude range             & $[55.843,\ 55.883]$ \\
\bottomrule
\end{tabular}
\end{table}

% ─── Section 3 ─────────────────────────────────────────────────────────────────
\section{Case Study: Modeling the Number of Daily Charging Sessions}
\label{Problemdes}

The main objective of this study is to model the number of charging 
sessions per day $t$ and per charging point $i$ in a spatio-temporal 
framework. While machine learning models can capture complex 
relationships by incorporating geographical coordinates and 
neighbourhood information as covariates, they offer limited 
interpretability in terms of spatial and temporal dependence 
structures.

We therefore adopt a statistical approach based on the Integrated 
Nested Laplace Approximation (INLA) methodology in \texttt{R}. INLA 
enables the formulation of hierarchical spatio-temporal models where 
spatial dependence is captured through Gaussian Markov Random Fields 
(GMRFs) and temporal dynamics are modelled explicitly. This provides 
an interpretable and probabilistic framework well suited to the 
structure of the data and to the goal of identifying drivers of EV 
charging demand across space and time.

\subsection{Model Description}

\begin{figure}[h!]
    \centering
    \includegraphics[width=0.4\linewidth]
    {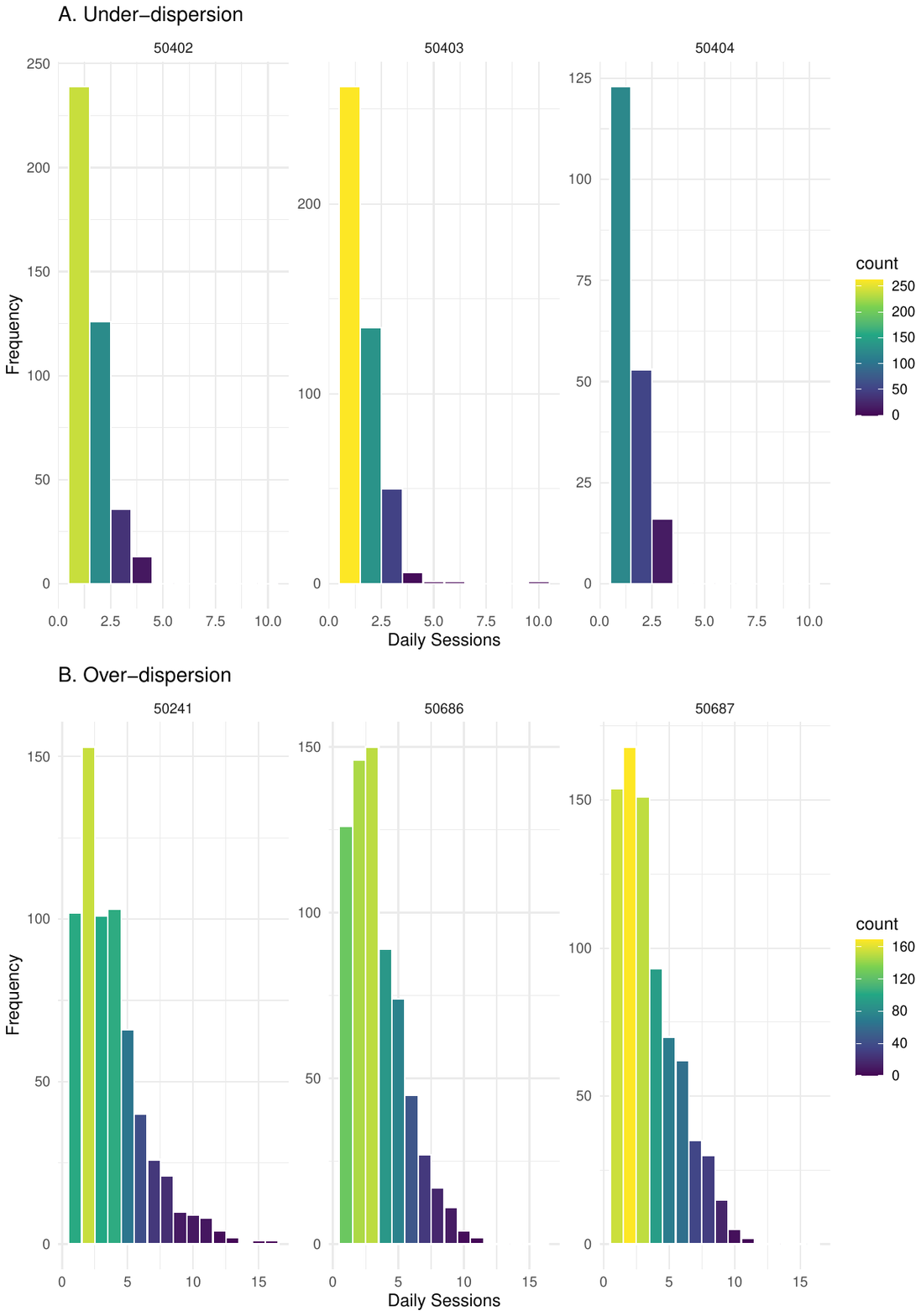}
    \caption{\small Empirical distributions of daily charging sessions 
    for six representative CPIDs. \textbf{(A)} CPIDs exhibiting 
    overdispersion. \textbf{(B)} CPIDs exhibiting underdispersion.}
    \label{dispersion_analysis}
\end{figure}

We model the number of daily charging sessions per charging point $i$ 
and day $t$ using a Poisson likelihood, which provides a natural 
baseline for count data. Figure~\ref{dispersion_analysis} shows the 
empirical distributions of daily session counts for a selection of 
CPIDs, revealing heterogeneity across units: some CPIDs exhibit 
overdispersion and others underdispersion. Despite this variability, 
the Poisson model offers an interpretable starting point that allows 
us to characterise spatial and temporal patterns before considering 
more flexible alternatives such as the Negative Binomial distribution.

\noindent\textbf{Models:} Let $y_{i,t}$ denote the number of charging 
sessions observed at charging point $i$ (CPID) on day $t$. We include 
covariates related to charging infrastructure (charger type, power 
capacity) and meteorological conditions (temperature, wind speed), 
which may influence usage patterns. Fixed effects capture these 
observed drivers, while structured random effects account for residual 
spatial and temporal heterogeneity.

\begin{equation}
y_{i,t} \mid \mu_{i,t} \sim \text{Poisson}(\mu_{i,t})
\end{equation}

where $\mu_{i,t}$ is the expected number of sessions for charging 
point $i$ on day $t$. We use a log link function, which is the 
canonical link for the Poisson distribution and ensures $\mu_{i,t} > 0$:

\begin{equation}
\mu_{i,t} = \exp(\eta_{i,t})
\end{equation}

The linear predictor $\eta_{i,t}$ decomposes into three components:

\begin{equation}
\eta_{i,t} = \alpha 
+ \sum_{k=1}^{K} \beta_k x_k(i,t) 
+ f_{\text{space}}(i) 
+ f_{\text{time}}(t)
\end{equation}

The fixed effects $\sum_{k=1}^{K} \beta_k x_k(i,t)$ capture the 
influence of observed covariates such as connector type and weather 
conditions. The spatial effect $f_{\text{space}}(i)$ captures 
unobserved heterogeneity across charging points, and the temporal 
effect $f_{\text{time}}(t)$ accounts for day-to-day dynamics and 
seasonal patterns shared across all units.

Both $f_{\text{space}}(i)$ and $f_{\text{time}}(t)$ are modelled as 
Gaussian Markov Random Fields (GMRFs), where the Markov property means 
that each effect depends only on its local neighbourhood; in time, 
this corresponds to adjacent days; in space, to geographically proximate 
charging points. This structure yields sparse precision matrices that 
enable efficient Bayesian inference via INLA. The following sections 
detail the specification of each component.

\subsubsection{Temporal Latent Dynamics: The Second-Order Random Walk}

The latent temporal effect $f_{\text{time}}(t)$ is modelled using a 
second-order Random Walk (RW2), which acts as a stochastic smoothing 
spline by penalising the curvature of the temporal trajectory 
\citep{RueHeld2005}. For $n_T$ days, the model assumes independent 
second-order differences:

\begin{equation}
\Delta^2 f_{\text{time}}(t) = f_{\text{time}}(t) - 2f_{\text{time}}(t+1) 
+ f_{\text{time}}(t+2) \sim \mathcal{N}(0, \tau_t^{-1}), 
\quad t = 1, \dots, n_T-2
\end{equation}

where $\tau_t$ is the precision parameter controlling smoothness. The 
joint prior is:

\begin{equation}
\pi(\mathbf{f} \mid \tau_t) \propto \tau_t^{\frac{n_T-2}{2}} 
\exp \left( -\frac{\tau_t}{2} \mathbf{f}^\top \mathbf{R} \mathbf{f} \right)
\end{equation}

This defines an Intrinsic GMRF with rank deficiency $k=2$, meaning the 
prior is invariant to global shifts and linear trends, which are 
absorbed by the fixed effects. The precision matrix $\mathbf{R}$ is 
sparse and banded, connecting each time point to its four nearest 
neighbours, which enables efficient inference via sparse Cholesky 
factorisation within INLA.

Inference is performed on the log-precision $\theta_t = \log \tau_t$, 
with a weakly informative prior:

\begin{equation}
\theta_t \sim \text{Log-Gamma}(a=1,\ b=5 \times 10^{-5})
\end{equation}

\subsubsection{Spatial Effect Modelling}

We consider two alternative specifications for the latent spatial 
effect $f_{\text{space}}(i)$.

\begin{enumerate}
    \item \textbf{Continuous approach (SPDE):} The spatial field is 
    modelled as a Gaussian Random Field (GRF) with Matérn covariance, 
    approximated by a GMRF defined on a triangulated mesh.

    \item \textbf{Discrete approach (ICAR):} The spatial domain is 
    partitioned into areal units and spatial dependence is captured 
    via an Intrinsic Conditional Autoregressive (ICAR) model, assuming 
    that neighbouring units have correlated latent effects.
\end{enumerate}

The SPDE approach captures smooth continuous variation over the study 
area, while ICAR is more interpretable in terms of neighbourhood-level 
correlations. The two strategies are described and compared in the 
following sections.

\noindent\textbf{Continuous Spatial Dynamics: The SPDE Approach}\\
\label{sec:spde_approach}

We model the latent spatial effect $f_{\text{spatial}}(s)$ as a 
Gaussian Random Field (GRF) with Matérn covariance. While GRFs provide 
a flexible framework for geostatistical modelling, exact inference 
requires inverting an $n \times n$ covariance matrix at $O(n^3)$ cost. 
Following \citep{LindgrenRueLindstrom2011}, we overcome this by representing 
$f_{\text{spatial}}(s)$ as the solution to a Stochastic Partial 
Differential Equation (SPDE):

\begin{equation}
(\kappa^2 - \Delta)^{\alpha/2} (\tau f_{\text{spatial}}(s)) = \mathcal{W}(s)
\end{equation}

where $\Delta$ is the Laplacian, $\kappa > 0$ is the scale parameter, 
$\alpha$ controls smoothness, and $\mathcal{W}(s)$ is spatial Gaussian 
white noise. We set $\alpha = 2$, corresponding to a Matérn smoothness 
$\nu = 1$ in $\mathbb{R}^2$, giving the covariance:

\begin{equation}
\text{Cov}(f_{\text{spatial}}(s_i), f_{\text{spatial}}(s_j)) = 
\sigma^2 \frac{2^{1-\nu}}{\Gamma(\nu)} 
(\kappa \|s_i - s_j\|)^\nu K_\nu(\kappa \|s_i - s_j\|)
\end{equation}

where $K_\nu$ is the modified Bessel function of the second kind. The 
field is fully characterised by the spatial range 
$\rho \approx \sqrt{8\nu}/\kappa$ and marginal variance $\sigma^2$:

\[
\sigma^2 = \frac{1}{4\pi\tau^2\kappa^2}, \qquad
\rho = \frac{\sqrt{8\nu}}{\kappa}
\]

Inference is performed on the log-transformed hyperparameters 
$\theta_1 = \log\tau$ and $\theta_2 = \log\kappa$, with weakly 
informative Gaussian priors:

\begin{equation}
\theta_1 \sim \mathcal{N}(\mu_\tau, \sigma^2_\tau), \qquad 
\theta_2 \sim \mathcal{N}(\mu_\kappa, \sigma^2_\kappa)
\end{equation}

The SPDE is solved numerically by projecting the field onto a 
triangulated mesh $\mathcal{T}$ via the Finite Element Method (FEM):

\begin{equation}
f_{\text{spatial}}(s) \approx \sum_{k=1}^{V} \psi_k(s) w_k
\end{equation}

This yields a GMRF with sparse precision matrix $\mathbf{Q}$, reducing 
computational complexity to $O(V^{3/2})$ \citep{Rue2009}.

The mesh is defined by three parameters. The inner edge, set to 200 m 
(one-fifth of the assumed interaction range of 1000 m, following 
\citep{Dambly2023}), controls the maximum edge length within the 
spatial domain. The outer edge, set to 2000 m following 
\citep{Righetto2020}, extends the mesh beyond the domain to avoid 
boundary effects. The cutoff, set to 100 m, defines the minimum 
distance between mesh nodes to avoid overly dense triangulation near 
observed points.

\begin{figure}[h!]
    
    \begin{minipage}{0.3\textwidth}
        \centering
        \resizebox{0.7\linewidth}{!}{
        $A = \begin{pmatrix}
        A_{11} & \cdots & A_{1V} \\
        \vdots & \ddots & \vdots \\
        A_{n1} & \cdots & A_{nV}
        \end{pmatrix}$
        }
        \vspace{6mm}
        {\footnotesize \textbf{(b) Projection matrix $\mathbf{A}$}}
    \end{minipage}
    \hfill
    \begin{minipage}{0.3\textwidth}
        \centering
        \scriptsize
        $f_{\text{spatial}}(s_i) = \sum_{k=1}^{3} A_{ik} f(v_k)$
        \vspace{2mm}
        \resizebox{0.75\linewidth}{!}{
            \begin{tikzpicture}
% Coordonnées
\coordinate (v3) at (0,0);
\coordinate (v2) at (6,0);
\coordinate (v1) at (3,4.6);
\coordinate (C)  at (3,1.85);

% Remplissage avec la couleur Viridis (opacité normale)
\fill[viridisBlue!80] (v3) -- (v2) -- (v1) -- cycle;

% Lignes de séparation internes (Blanches)
\draw[white, line width=1.4pt] (C) -- (v1);
\draw[white, line width=1.4pt] (C) -- (v2);
\draw[white, line width=1.4pt] (C) -- (v3);

% Bordure extérieure (couleur pleine)
\draw[viridisBlue, line width=0.8pt] (v3) -- (v2) -- (v1) -- cycle;

% Nodes des Aires (Texte en blanc pour contraste sur le bleu)
\node[white, font=\bfseries\small] at (3.7,2.45) {A1};
\node[white, font=\bfseries\small] at (3.0,1.05) {A2};
\node[white, font=\bfseries\small] at (2.3,2.45) {A3};

% Nodes des Sommets (Texte en couleur Viridis pour rappel)
\node[viridisBlue, font=\bfseries] at (v1) [above] {v1};
\node[viridisBlue, font=\bfseries] at (v2) [below right] {v2};
\node[viridisBlue, font=\bfseries] at (v3) [below left] {v3};

\end{tikzpicture}
        }
        \vspace{2mm}
        {\footnotesize \textbf{(c) Barycentric interpolation}}
    \end{minipage}

    \caption{\small SPDE numerical workflow:  
    (b) projection matrix $\mathbf{A}$ mapping mesh nodes to observation 
    locations, and (c) barycentric interpolation on the triangular 
    manifold.}
    \label{fig:spatial_mapping}
\end{figure}
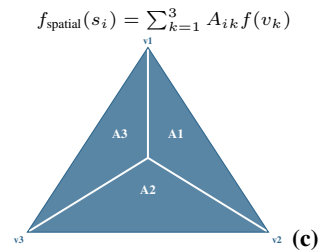

Observed charging locations $s_i$ are linked to the mesh via a sparse 
projection matrix $\mathbf{A}$. For a location $s_i$ within a triangle 
with vertices $\{v_1, v_2, v_3\}$, the field is interpolated as:

\begin{equation}
f_{\text{spatial}}(s_i) = \sum_{k=1}^{3} A_{ik} f(v_k)
\end{equation}

where $A_{ik}$ are the barycentric coordinates at $s_i$. This workflow 
is illustrated in Figure~\ref{fig:spatial_mapping}.

\begin{figure}
    \centering
 \includegraphics[width=0.5\linewidth, trim=2mm 1.5mm 1.5mm 2mm, 
        clip]{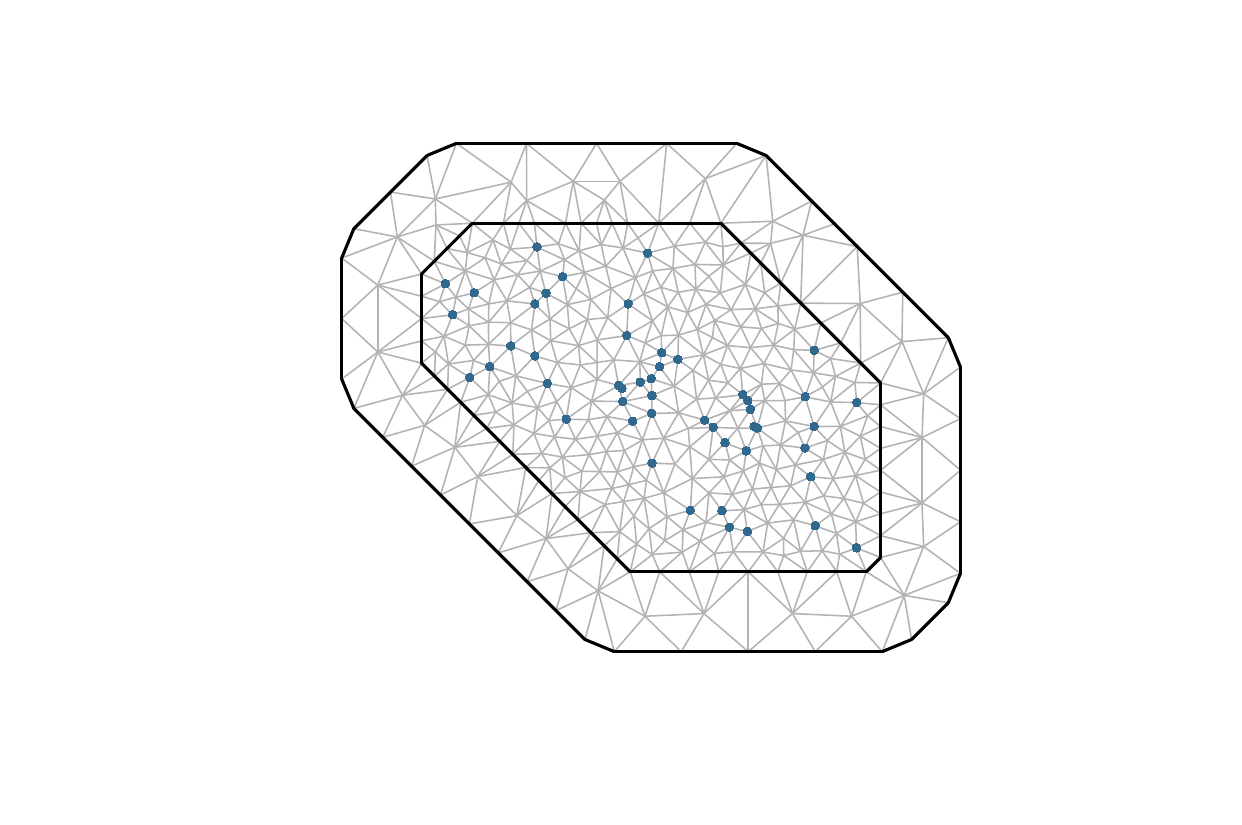}
    \caption{\small Mesh construction via Constrained 
        Delaunay Triangulation}
    \label{fig:placeholder}

\end{figure}
\noindent\textbf{Discrete Spatial Dynamics: The ICAR Specification}
\label{sec:icar_approach}

As an alternative to the SPDE approach, we implement a discrete spatial 
model using the Intrinsic Conditional AutoRegressive (ICAR) 
specification \citep{Besag1974}. The study area is partitioned into 
areal units (CPIDs), and the latent spatial effect 
$f_{\text{spatial}}(s_i)$ is modelled conditionally as:

\begin{equation}
f_{\text{spatial}}(s_i) \mid f_{\text{spatial}}(s_{-i}) \sim 
\mathcal{N}\left( \frac{1}{n_i} \sum_{j \in \delta_i} 
f_{\text{spatial}}(s_j),\ \frac{\sigma^2}{n_i} \right)
\end{equation}

where $n_i$ is the number of neighbours of unit $i$, $\delta_i$ is the 
set of their indices, and $\sigma^2$ controls the magnitude of spatial 
variation. The joint distribution is a GMRF with precision matrix 
$\mathbf{Q} = \tau(\mathbf{D} - \mathbf{W})$, where $\mathbf{W}$ is 
the adjacency matrix and $\mathbf{D}$ is the diagonal degree matrix. 
Model identifiability is ensured by a sum-to-zero constraint:

\begin{equation}
\sum_{i=1}^{n} f_{\text{spatial}}(s_i) = 0
\end{equation}

\begin{figure}[h]
    \centering
    \includegraphics[width=0.4\linewidth]
    {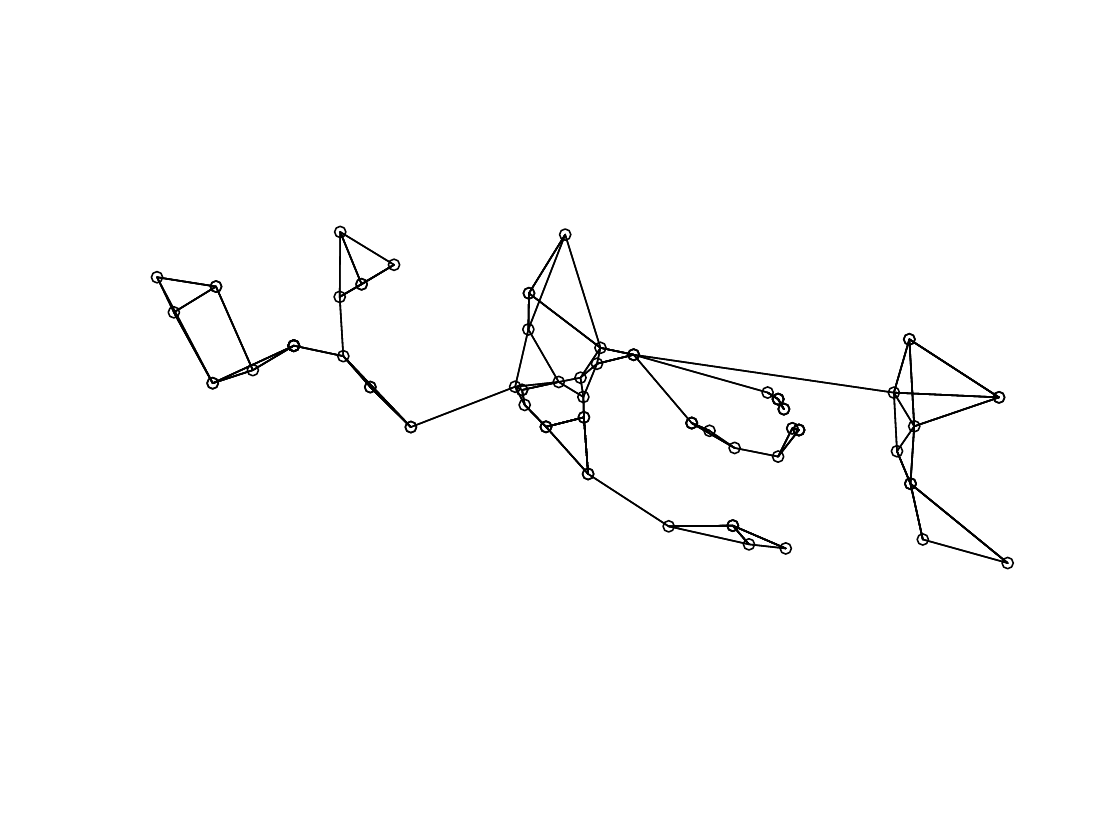}
    \caption{\small Spatial connectivity graph of the Glasgow EV 
    charging network. Nodes represent CPIDs positioned by their 
    geographical coordinates. Edges define the adjacency structure 
    for the ICAR model, constructed using $k$-nearest neighbours 
    ($k=4$) with bridging edges to ensure a single connected component.}
    \label{fig:icar_graph}
\end{figure}

\noindent\textbf{Neighbourhood Structure.}
Neighbours are defined using a $k$-nearest neighbours approach ($k=4$) 
based on the geographical coordinates of the CPIDs. To ensure the 
adjacency graph is connected, isolated nodes are manually linked to 
their nearest neighbour, guaranteeing a well-defined precision matrix 
and a unique sum-to-zero constraint. The resulting graph is shown in 
Figure~\ref{fig:icar_graph}.

Table~\ref{tab:model_components} summarises all model components, 
their prior distributions, and interpretations.

\begin{table}[h!]
\centering
\footnotesize
\caption{Summary of model components, prior distributions, and 
interpretations.}
\label{tab:model_components}
\renewcommand{\arraystretch}{1.3}
\begin{tabularx}{0.85\textwidth}{l >{\raggedright\arraybackslash}p{3.8cm} 
>{\raggedright\arraybackslash}X}
\toprule
\textbf{Component} & \textbf{Prior} & \textbf{Interpretation} \\
\midrule
Fixed effects $\beta_k$ & 
Gaussian, weakly informative & 
Effect of covariates (charger type, weather, etc.) \\
Temporal effect $f_{\text{time}}(t)$ & 
RW2: $\Delta^2 f_t \sim \mathcal{N}(0, \tau_t^{-1})$ & 
Smooth non-stationary temporal trend \\
Temporal precision $\tau_t$ & 
Log-Gamma($a=1,\ b=5\times10^{-5}$) & 
Controls RW2 smoothness \\
Spatial effect $f_{\text{spatial}}(s)$ & 
SPDE/GMRF (Matérn) & 
Smooth continuous spatial variation \\
SPDE precision $\tau$ & 
Log-Gaussian & 
Marginal variance of the spatial field \\
SPDE scale $\kappa$ & 
Log-Gaussian & 
Spatial correlation range $\rho$ \\
$f_{\text{spatial}}(s_i)$ & 
$\mathcal{N}\!\left( \frac{1}{n_i} \sum_{j \in \delta_i} 
f_{\text{spatial}}(s_j), \frac{\sigma^2}{n_i} \right)$ & 
Neighbourhood-level spatial smoothing \\
ICAR variance $\sigma^2$ & 
Weakly informative & 
Magnitude of spatial variability \\
\bottomrule
\end{tabularx}
\end{table}

\begin{figure}[h]
    \centering
    \includegraphics[width=0.7\linewidth]
    {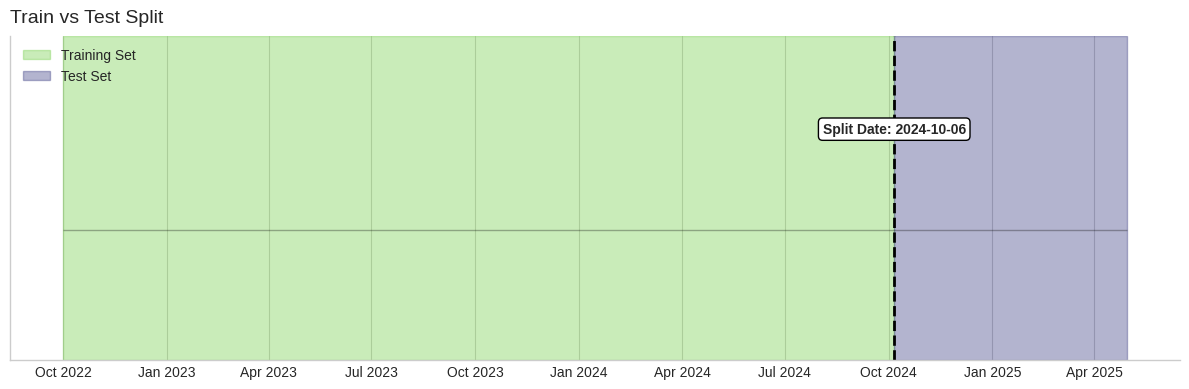}
    \caption{\small Temporal train--test split of the dataset.}
    \label{fig:data_split}
\end{figure}

\noindent\textbf{Data Splitting.}
The dataset is split into a training set ($\mathcal{D}_{\text{train}}$) 
and a test set ($\mathcal{D}_{\text{test}}$) following an 80/20 
temporal split, with the split date at 2024-10-06. The training set 
is used for parameter estimation and the test set provides an 
out-of-sample evaluation window. Figure~\ref{fig:data_split} 
illustrates this partition.

\noindent\textbf{Evaluation.}
Model performance is assessed on $\mathcal{D}_{\text{test}}$ using 
three metrics: Mean Absolute Error (MAE), Root Mean Squared Error 
(RMSE), and Mean Absolute Percentage Error (MAPE). MAE measures 
average prediction error, RMSE penalises larger deviations, and MAPE 
provides a scale-independent measure that allows comparison across 
CPIDs with different demand levels.

\noindent\textbf{Benchmarking Strategy.}
The spatio-temporal models are compared against two station-level 
baselines: a Poisson GLM and XGBoost, both fitted independently for 
each CPID without any spatial information sharing. This comparison 
allows us to quantify the benefit of pooling information across the 
network through structured spatial priors.

% ─── Section 4 ─────────────────────────────────────────────────────────────────
\section{Empirical Results}
\label{results}
 
We fit two hierarchical spatio-temporal models using INLA: one with a 
continuous spatial component (SPDE--RW2) and one with a discrete 
spatial component (ICAR--RW2). Both models include a second-order 
Random Walk (RW2) for temporal dynamics and the same set of fixed 
effects. Model fit is assessed using WAIC and DIC, and out-of-sample 
predictive performance is evaluated on $\mathcal{D}_{\text{test}}$ 
using MAE, RMSE, and MAPE, benchmarked against station-level Poisson 
GLM and XGBoost models.

\noindent\textbf{Model Fit.}
The ICAR--RW2 model achieves a slightly better fit than SPDE--RW2 
according to both information criteria (ICAR: DIC~$= 136{,}562.8$, 
WAIC~$= 136{,}509.1$; SPDE: DIC~$= 136{,}941.3$, 
WAIC~$= 136{,}909.1$). The posterior estimates are consistent across 
both specifications, suggesting that the main drivers of demand are 
robust to the choice of spatial structure.

\begin{table}[h!]
\centering
\footnotesize
\caption{Posterior estimates of fixed effects for the ICAR--RW2 
and SPDE--RW2 models.}
\label{tab:fixed_effects}
\begin{tabular}{lcccc}
\toprule
& \multicolumn{2}{c}{\textbf{ICAR--RW2}} 
& \multicolumn{2}{c}{\textbf{SPDE--RW2}} \\
\cmidrule(lr){2-3} \cmidrule(lr){4-5}
\textbf{Covariate} & Mean & 95\% CI & Mean & 95\% CI \\
\midrule
\multicolumn{5}{l}{\textit{Infrastructure}} \\
Intercept          & 0.592 & $[0.344, 0.841]$ & 0.331 & $[0.228, 0.438]$ \\
Rapid connector    & 0.872 & $[0.599, 1.146]$ & 0.914 & $[0.819, 1.008]$ \\
Public access      & 0.321 & $[0.107, 0.534]$ & 0.350 & $[0.258, 0.443]$ \\
Free charging      & 0.146 & $[0.120, 0.173]$ & 0.142 & $[0.116, 0.169]$ \\
\midrule
\multicolumn{5}{l}{\textit{Day of week (ref: Friday)}} \\
Monday             & $-$0.019 & $[-0.042,\ 0.003]$ & $-$0.020 & $[-0.043,\ 0.002]$ \\
Tuesday            & $-$0.016 & $[-0.038,\ 0.007]$ & $-$0.017 & $[-0.039,\ 0.006]$ \\
Wednesday          & 0.002  & $[-0.020,\ 0.025]$ & 0.003  & $[-0.019,\ 0.025]$ \\
Thursday           & $-$0.004 & $[-0.026,\ 0.018]$ & $-$0.005 & $[-0.027,\ 0.017]$ \\
Saturday           & $-$0.030 & $[-0.053,\ -0.007]$ & $-$0.029 & $[-0.053,\ -0.006]$ \\
Sunday             & $-$0.063 & $[-0.087,\ -0.040]$ & $-$0.066 & $[-0.090,\ -0.042]$ \\
\midrule
\multicolumn{5}{l}{\textit{Weather}} \\
Temp.\ spline 1    & $-$0.324 & $[-0.481,\ -0.168]$ & $-$0.363 & $[-0.517,\ -0.209]$ \\
Temp.\ spline 2    & $-$0.011 & $[-0.242,\ 0.220]$  & $-$0.243 & $[-0.327,\ -0.158]$ \\
Temp.\ spline 3    & 0.046  & $[-0.258,\ 0.349]$  & $-$0.294 & $[-0.409,\ -0.178]$ \\
\midrule
\multicolumn{5}{l}{\textit{Model fit}} \\
DIC  & \multicolumn{2}{c}{136,562.8} & \multicolumn{2}{c}{136,941.3} \\
WAIC & \multicolumn{2}{c}{136,509.1} & \multicolumn{2}{c}{136,909.1} \\
\bottomrule
\end{tabular}
\end{table}

\begin{table}[h!]
\centering
\footnotesize
\caption{Posterior estimates of model hyperparameters.}
\label{tab:hyperparameters}
\begin{tabular}{lccc}
\toprule
\textbf{Parameter} & \textbf{Mean} & \textbf{SD} & \textbf{95\% CI} \\
\midrule
\multicolumn{4}{l}{\textit{ICAR--RW2}} \\
Precision (ICAR) $\tau$ & 2.54 & 0.39 & $[1.84,\ 3.38]$ \\
Precision (RW2) $\tau_t$ & $1.99\times10^5$ & $4.61\times10^4$ 
                          & $[1.22\times10^5,\ 3.03\times10^5]$ \\
\midrule
\multicolumn{4}{l}{\textit{SPDE--RW2}} \\
$\theta_1 = \log\tau$   & 5.16    & 0.30 & $[4.53,\ 5.72]$ \\
$\theta_2 = \log\kappa$ & $-$5.44 & 0.32 & $[-6.03,\ -4.78]$ \\
Precision (RW2) $\tau_t$ & $1.97\times10^5$ & $4.62\times10^4$ 
                          & $[1.21\times10^5,\ 3.02\times10^5]$ \\
\bottomrule
\end{tabular}
\end{table}

\noindent\textbf{Fixed Effects.}
Both models agree on the main drivers of demand. Rapid connectors 
have the strongest positive effect (ICAR: $\hat{\beta} = 0.872$, 
95\%~CI~$[0.599, 1.146]$; SPDE: $\hat{\beta} = 0.914$, 
95\%~CI~$[0.819, 1.008]$), indicating that Rapid charge points 
attract more daily sessions than AC ones. Public access and 
free-of-charge status also have positive and consistent effects 
across both models (ICAR: $\hat{\beta}_{\text{free}} = 0.146$; 
SPDE: $\hat{\beta}_{\text{free}} = 0.142$). Among weather 
variables, only the first temperature spline shows a credible 
negative effect, while humidity and wind speed have negligible 
influence. Regarding day-of-week, Sunday records the largest 
negative coefficient (ICAR: $\hat{\beta} = -0.063$; 
SPDE: $\hat{\beta} = -0.066$), consistent with the lower weekend 
demand observed in the EDA.

\begin{figure}[h!]
    \centering
    % --- Première rangée : Comparaison des Modèles ---
    \begin{subfigure}[b]{0.4\textwidth}
        \centering
        \includegraphics[width=\textwidth, trim={0 0 0 1.2cm}, clip]{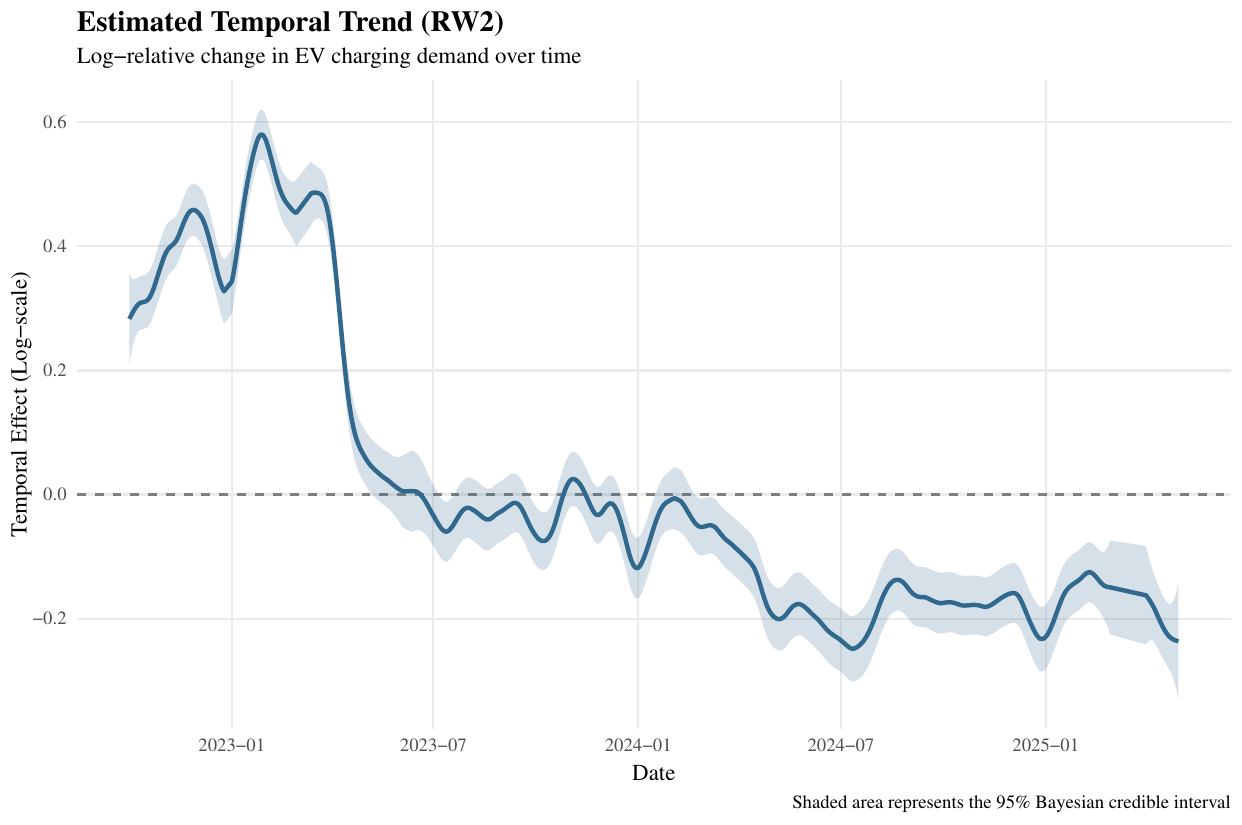}
        \caption{\small SPDE--RW2 Latent Trend}
        \label{fig:spde_rw2}
    \end{subfigure}
    \hfill
    \begin{subfigure}[b]{0.4\textwidth}
        \centering
        \includegraphics[width=\textwidth, trim={0 0 0 1.2cm}, clip]{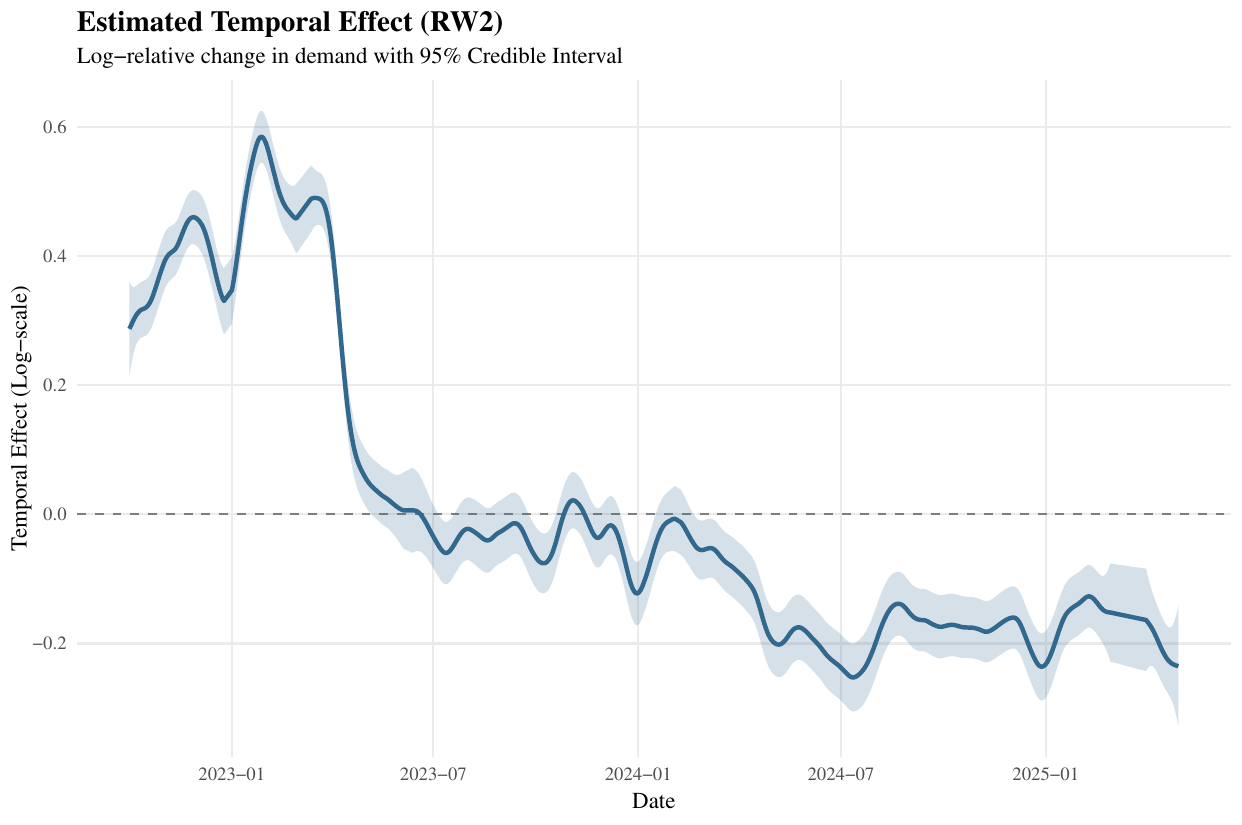}
        \caption{\small ICAR--RW2 Latent Trend}
        \label{fig:icar_rw2}
    \end{subfigure}

    \vspace{2mm}
    \begin{subfigure}[b]{0.4\textwidth} 
        \centering
        \includegraphics[width=\textwidth, trim={0 0 0 1.25cm}, clip]{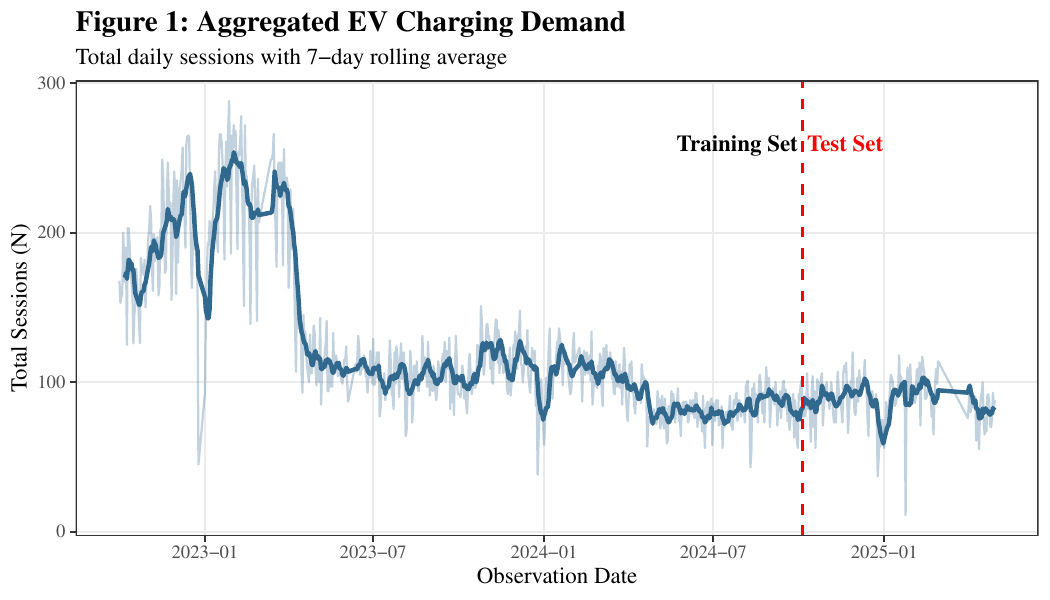}
        \caption{\small Observed Global EV Demand}
        \label{fig:global_demand}
    \end{subfigure}

    \vspace{2mm}
    \caption{\small \textbf{Temporal dynamics and structural break analysis.} Panels \textbf{(a)} and \textbf{(b)} display the posterior mean of the RW2 latent temporal component. Panel \textbf{(c)} presents the aggregated observed EV charging demand.}
    \label{fig:time_effects}
\end{figure}

\noindent\textbf{Temporal Dynamics. }Figure~\ref{fig:time_effects} shows the posterior mean of the RW2 
latent temporal component for both models. The estimated trend captures 
two main features: a sharp decline in demand following the introduction 
of tariffs in early 2023, and a gradual stabilisation thereafter. The 
high posterior precision of the RW2 effect 
($\tau \approx 1.96 \times 10^5$) indicates that the temporal 
specification absorbs most of the residual time-varying signal, 
leaving little unexplained temporal variance.

\begin{figure}[h!]
    \centering
    \begin{subfigure}[b]{0.4\textwidth}
        \centering
        \includegraphics[width=\textwidth]
        {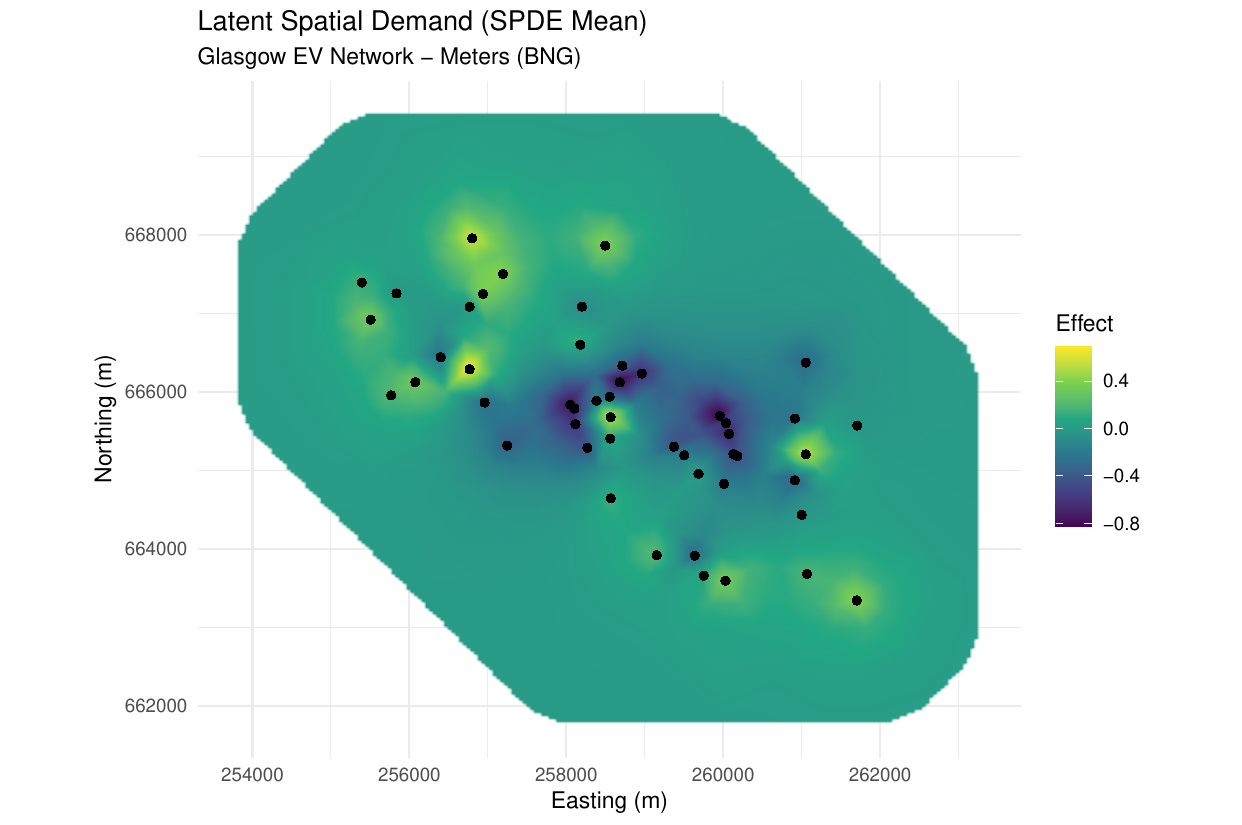}
        \caption{\small SPDE (Continuous)}
        \label{fig:spde_mean}
    \end{subfigure}
    \hfill
    \begin{subfigure}[b]{0.4\textwidth}
        \centering
        \includegraphics[width=\textwidth]
        {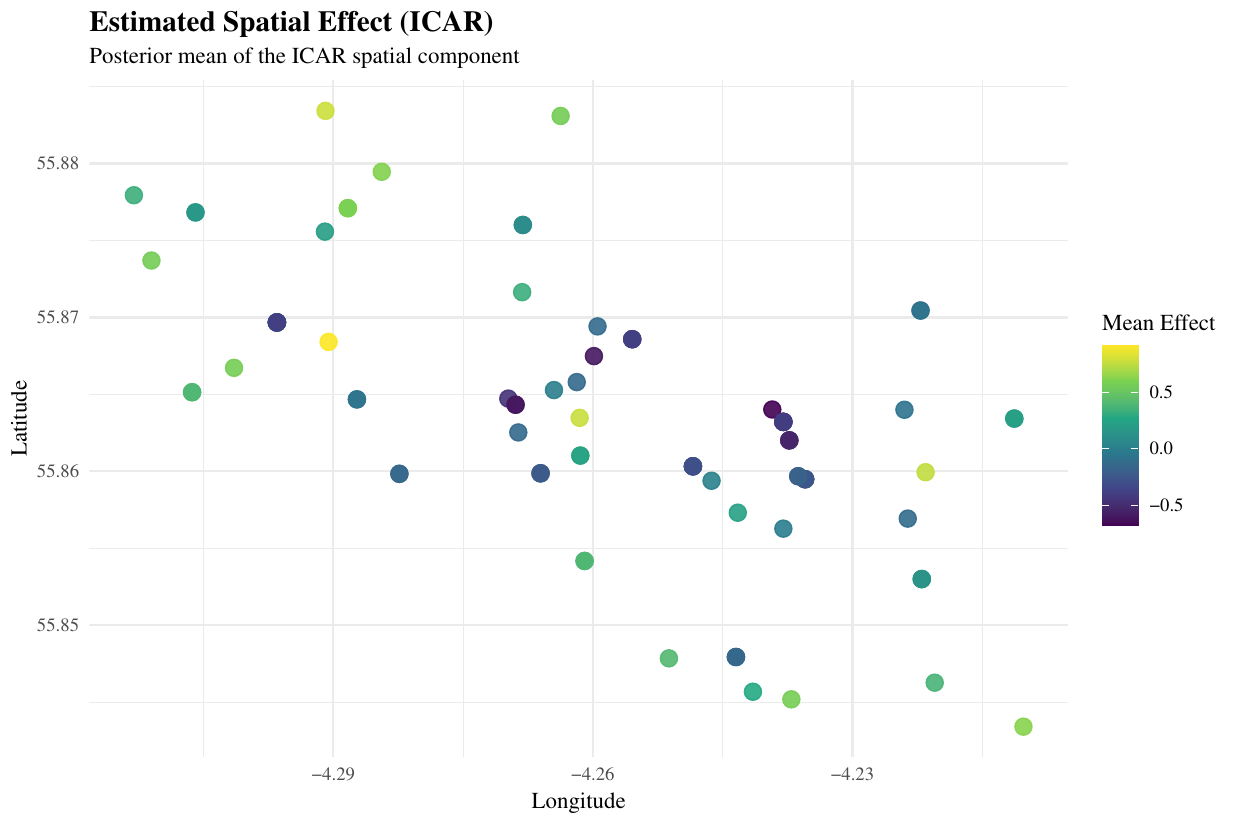}
        \caption{\small ICAR (Discrete)}
        \label{fig:icar_mean}
    \end{subfigure}
    \caption{\small Posterior mean of the latent spatial effect for 
    the SPDE (a) and ICAR (b) specifications.}
    \label{fig:spatial_comparison}
\end{figure}

\noindent\textbf{Spatial Correlation.}
The SPDE model estimates a posterior mean spatial range of 
$\hat\rho = 684.4$~m (95\%~CI: $[338.9, 1175.1]$~m). Given that the 
mean nearest-neighbour distance between CPIDs is 113.8~m and the 
maximum is 834.5~m, this range is physically coherent: it captures 
interactions between nearby clusters of chargers without over-smoothing 
across distinct neighbourhoods. This suggests that spatial dependencies 
in EV charging demand operate at a sub-kilometre scale, likely driven 
by local land use and amenity patterns. Figure~\ref{fig:spatial_comparison} shows the posterior mean spatial 
effects from both models. The two specifications recover similar 
spatial patterns: positive effects (yellow) correspond to CPIDs with 
higher demand relative to their neighbourhood average, while negative 
effects (blue) indicate lower-than-average activity. Notably, 
high-density areas tend to distribute load across multiple CPIDs, 
whereas isolated CPIDs in low-density zones carry proportionally 
higher demand.

\begin{figure}[h!]
    \centering
    % --- Comparaison des trois métriques sur une seule ligne ---
    \includegraphics[width=0.6\linewidth]{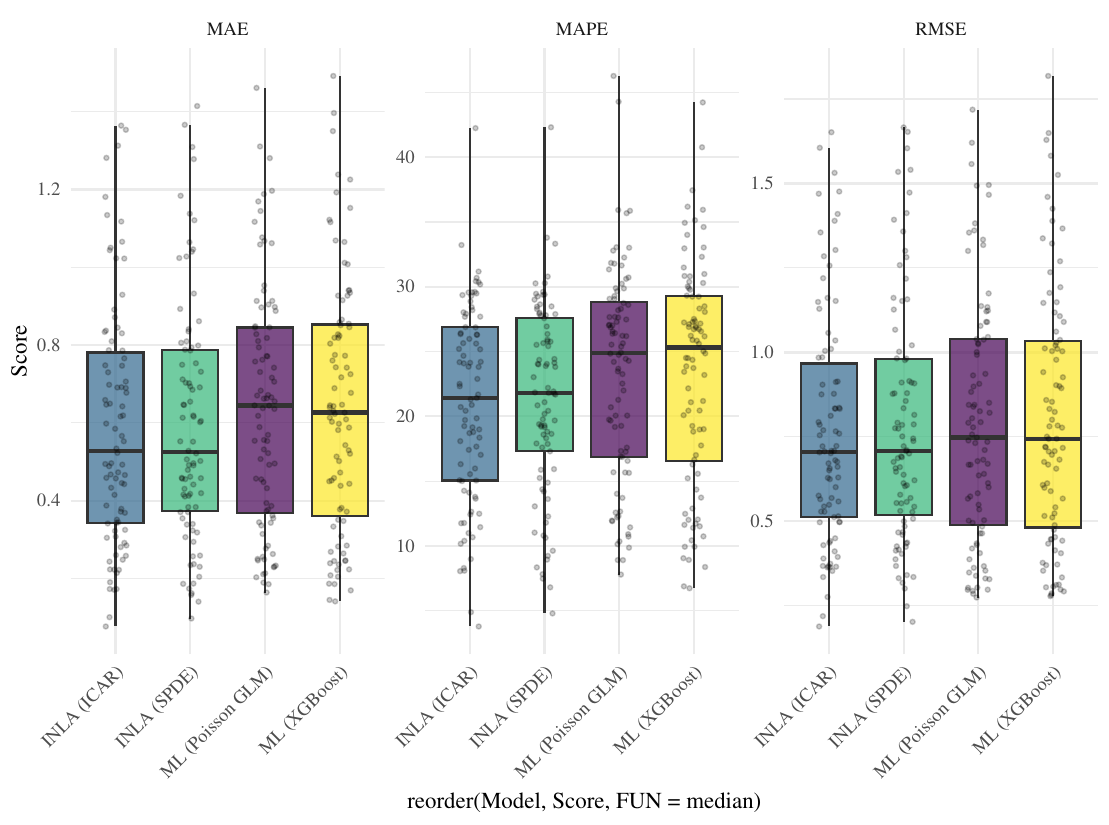}

    \caption{\small \textbf{Predictive performance benchmarking.} Comparison of (a) Root Mean Squared Error, (b) Mean Absolute Error, and (c) Mean Absolute Percentage Error across three model classes: 
    \textit{(i) Local Regressors} (XGBoost, GAM), 
    \textit{(ii) Bayesian Spatio-Temporal Models} (ICAR--RW2, SPDE--RW2)}
    \label{fig:model_performance}
\end{figure}

\newpage
\noindent\textbf{Benchmark of Spatio-Temporal and Local Models.} Figure~\ref{fig:model_performance} shows the distribution of 
station-level errors across all models. On MAE, the INLA models 
show slightly lower medians and more concentrated distributions 
than the station-level baselines, suggesting more consistent 
predictions across the network. On MAPE, the advantage is more 
visible: INLA distributions are shifted downward with fewer extreme 
values, indicating better relative accuracy especially for 
low-demand stations. On RMSE, the four models are difficult to 
distinguish; the distributions overlap substantially, suggesting 
near-parity on this metric.

\begin{figure}[h!]
    \centering
    
    \begin{subfigure}{0.31\textwidth}
        \includegraphics[width=\linewidth]{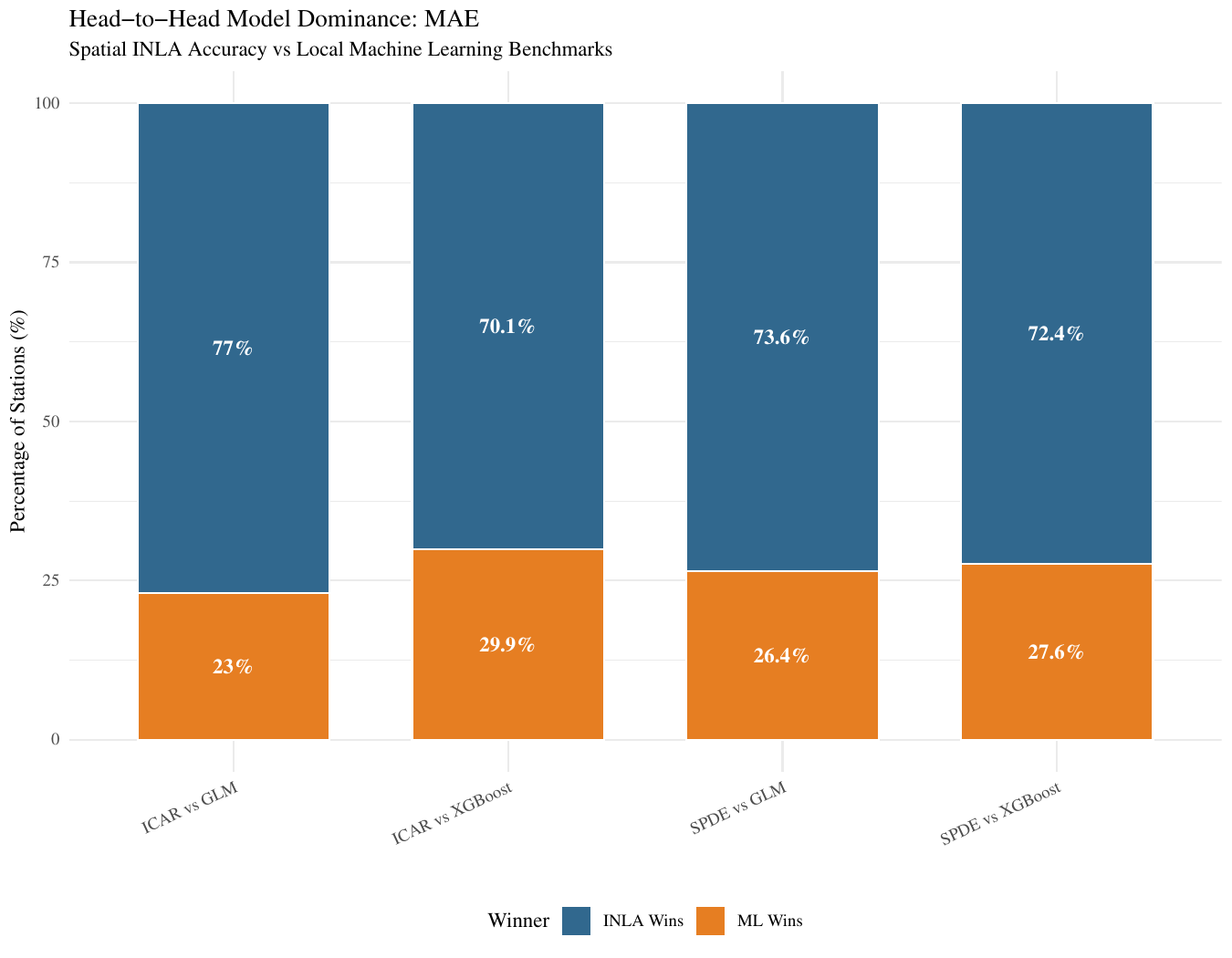}
        \caption{MAE}
    \end{subfigure}
    \hfill
    \begin{subfigure}{0.31\textwidth}
        \includegraphics[width=\linewidth]{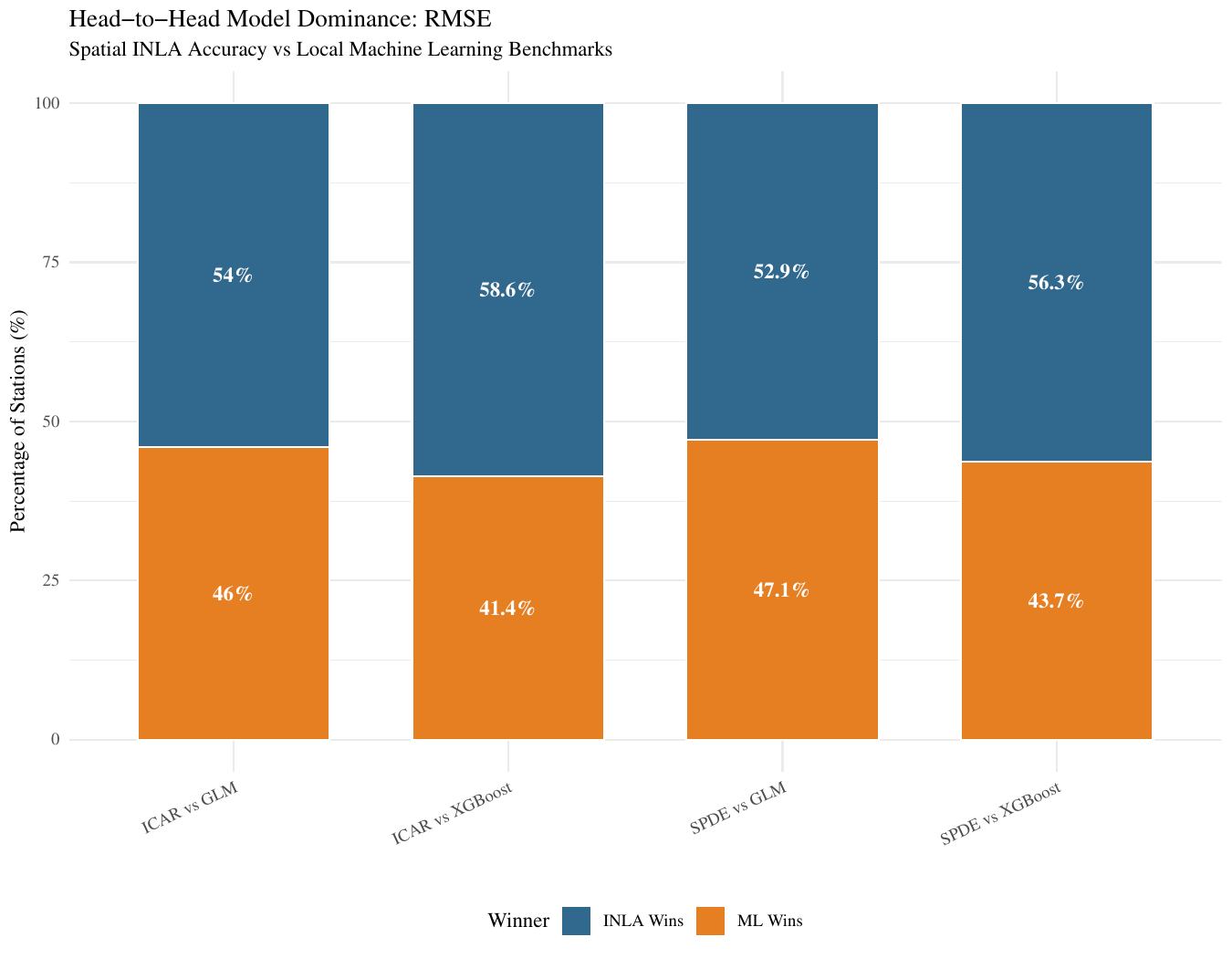}
        \caption{RMSE}
    \end{subfigure}
    \hfill
    \begin{subfigure}{0.31\textwidth}
        \includegraphics[width=\linewidth]{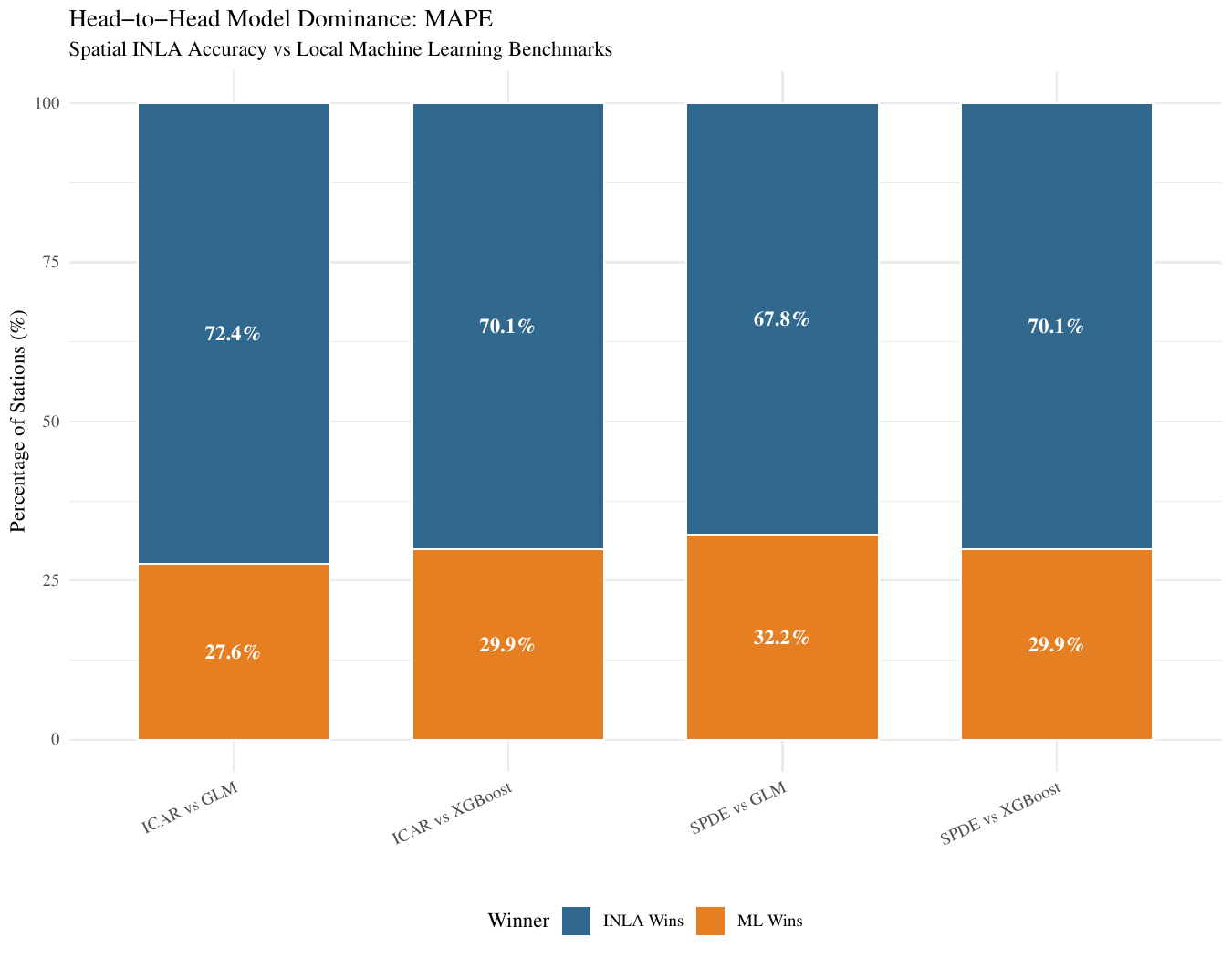}
        \caption{MAPE}
    \end{subfigure}
    
    \caption{\small Percentage of stations where INLA outperforms ML models across error metrics.}
    \label{pourcentage}
\end{figure}

Figure~\ref{pourcentage} confirms these observations at the 
station level. On MAE, ICAR--RW2 leads XGBoost at $77\%$ of 
stations and SPDE--RW2 at $73.6\%$, with both models achieving 
$70$--$72\%$ dominance against the Poisson GLM. Results on MAPE 
follow a similar pattern, with INLA leading at $68$--$72\%$ of 
stations. On RMSE, the dominance rates drop to $53$--$59\%$, 
confirming near-parity with the baselines. This difference between 
MAE/MAPE and RMSE suggests that spatial pooling reduces average 
and relative errors consistently, but does not systematically 
reduce large individual errors, which are likely driven by highly 
local demand spikes not captured by the spatial structure.

% ─── Section 5 ─────────────────────────────────────────────────────────────────
\section{Summary of Findings and Discussion}
 
This paper makes two contributions. First, we introduce a 
longitudinal dataset of EV charging sessions in Scotland, covering 
the period from October 2022 to April 2025. The dataset is sourced 
from the ChargePlace Scotland open-access repository and enriched 
with geographic metadata and meteorological covariates. A 
reproducible preprocessing pipeline is provided, covering data 
cleaning, standardisation, and spatial indexing. To our knowledge, 
this is one of the few large-scale, station-level EV charging 
datasets covering a recent period that captures the current phase 
of EV. Second, we use this dataset as a 
benchmark to evaluate a spatio-temporal modelling framework based 
on INLA, comparing it against station-level machine learning 
baselines.

\noindent\textbf{Key findings on the data.}
The exploratory analysis reveals several patterns that are relevant 
both for modelling and for infrastructure planning. The introduction 
of tariffs between 2022 and 2025 had a strong and lasting impact on 
demand: regions that transitioned from free to paid access saw sharp 
declines in session volumes, while regions with pre-existing tariffs 
remained stable. This confirms that charging status must be treated 
as a time-varying covariate in any predictive model applied to this 
dataset. Rapid chargers are used on $83\%$ of days on average, 
compared to $59\%$ for AC chargers, and attract more than twice the 
mean daily sessions (4.87 vs 2.22). Demand follows clear weekly 
cycles, with Sundays consistently recording the lowest activity. 
These patterns are stable across the observation period and are 
well captured by the fixed effects in both INLA models.

\noindent\textbf{Key findings on the modelling.}
The INLA framework captures the main spatial and temporal structure 
of the data. The estimated spatial range of $\hat{\rho} \approx 684$ 
m is physically coherent with the Glasgow network, indicating that 
spatial dependencies operate at a sub-kilometre scale, likely driven 
by local land use and amenity patterns. The RW2 latent trend 
successfully tracks the long-term evolution of demand, including the 
structural break caused by tariff introduction in early 2023. In 
terms of predictive performance, the INLA models are competitive 
with station-level ML baselines on MAE and MAPE, with dominance 
rates of $70$--$77\%$ against XGBoost and the Poisson GLM. On RMSE, 
the two approaches perform at a similar level, suggesting that 
spatial pooling reduces average errors but does not systematically 
reduce large individual prediction errors. The main added value of 
the INLA framework over station-level models lies in its 
interpretable decomposition of spatial and temporal effects and its 
principled uncertainty quantification, which are difficult to obtain 
from black-box ML models.

\noindent\textbf{Limitations.}
Several limitations should be noted. The modelling case study focuses 
on central Glasgow, which was selected for its high session density. 
Results may not generalise directly to sparser regions of the 
Scottish network where fewer observations per CPID are available. 
The Poisson likelihood, while a natural baseline for count data, is 
imperfect given the observed mix of over- and underdispersion across 
CPIDs. More flexible likelihoods such as the Negative Binomial 
could be explored in future work. Finally, the dataset covers 
sessions recorded by the ChargePlace Scotland network, which may 
not be representative of private or workplace charging behaviour 
in Scotland as a whole.

% ─── Section 6 ─────────────────────────────────────────────────────────────────
\section{Future Directions}
 
This work opens several directions for future research, both from a 
methodological and an applied perspective.

\noindent\textbf{Scaling to Scotland.}
The current study focuses on central Glasgow, selected for its high 
session density and network complexity. A natural next step is to 
extend the analysis to the full Scottish network, which would require 
handling a much sparser and more heterogeneous spatial distribution 
of charge points. This raises non-trivial modelling questions around 
the choice of neighbourhood structure, the treatment of regions with 
very few CPIDs, and the stability of spatial estimates in low-density 
areas.

\noindent\textbf{Hierarchical Forecasting.}
The current model operates at the CPID level. Extending it to a 
hierarchical framework would allow predictions at multiple spatial 
scales simultaneously; individual charge points, neighbourhoods, 
and local authorities; while ensuring coherence across aggregation 
levels. This is particularly relevant for grid operators who need 
consistent forecasts at both fine and coarse spatial resolutions.

\noindent\textbf{Modelling Other Variables of Interest.}
The present work focuses on the number of daily sessions. Future 
work could extend the framework to model energy load profiles and 
peak power consumption, which are more directly relevant for grid 
management. These variables present additional modelling challenges, 
such as heavier-tailed distributions and stronger temporal 
dependencies, that would require appropriate likelihood specifications 
beyond the Poisson model.

\noindent\textbf{Hybrid INLA--ML Approach.}
This study treats INLA and machine learning as competing approaches. 
A more promising direction is to combine them in a two-stage 
pipeline. In the first stage, a machine learning model; such as 
XGBoost;captures complex non-linear covariate interactions that 
are difficult to encode within the GMRF framework. In the second 
stage, INLA models the residuals of the first stage, which typically 
exhibit lower variance and a more linear structure than the raw 
demand signal \citep{hu2026xgboostmeetsinlatwostage}. This decomposition would allow the 
hybrid model to benefit from the flexibility of machine learning and 
the principled uncertainty quantification of Bayesian inference 
simultaneously.

% ───────────────────────────────────────────────────────────────────────────────
%  SUPPLEMENTARY MATERIALS
% ───────────────────────────────────────────────────────────────────────────────
\section*{Supplementary Materials}
 
All the codes used in this study are available at the following Git repository:
\href{https://gitlab.com/kaoutarbouaachra/Spatio-Temporal-Modelling-of-Electric-Vehicle-Charging-Demand}{Git repository}.
 
The datasets supporting the findings of this study are publicly available on Zenodo:
\href{https://zenodo.org/records/18727698}{Zenodo repository}.
The Zenodo repository contains the processed EV charging station usage data,
including temporal, spatial, and technical covariates used in the analysis.
 
% ───────────────────────────────────────────────────────────────────────────────
%  BROADER IMPACT
% ───────────────────────────────────────────────────────────────────────────────
\section*{Broader Impact}
 
This work develops a spatio-temporal forecasting model for electric vehicle (EV)
charging demand using publicly available data. The dataset is fully open and does
not contain any personal or identifiable information about individual users or
vehicles. Therefore, to the best of our knowledge, this research does not pose
risks related to privacy, consent, security, or human rights.
 
On the positive side, improved forecasting of EV charging demand can support
better infrastructure planning, grid management, and integration of renewable
energy sources. This may contribute to more efficient energy systems and
facilitate the transition toward sustainable transportation.
 
No foreseeable harmful societal impacts are associated with this work.
 
% ───────────────────────────────────────────────────────────────────────────────
%  ACKNOWLEDGEMENTS
% ───────────────────────────────────────────────────────────────────────────────
\section*{Acknowledgements}
 
The authors gratefully acknowledge ChargePlace Scotland for providing open
access to the electric vehicle charging dataset used in this study, which made
this research possible. This work received no specific grant from any funding
agency in the public, commercial, or not-for-profit sectors. The authors declare
that they have no competing financial interests or personal relationships that
could have influenced the work reported in this paper.
 
\bibliographystyle{unsrt}
\bibliography{sample} 
\end{document}